\documentclass[usenatbib]{mn2e}
\usepackage{astrojournals}
\usepackage{graphicx} 
\usepackage{amsmath}
\usepackage{times}
\usepackage{url}
\usepackage{amssymb}

\newcommand{\beq}{\begin{equation}}
\newcommand{\eeq}{\end{equation}}

\newcommand{\Rvir}{r_{\mathrm{vir}}}
\newcommand{\Mvir}{M_{\mathrm{vir}}}

\newcommand{\Rmax}{r_{\mathrm{max}}}
\newcommand{\Vmax}{V_{\mathrm{max}}}
\newcommand{\rs}{r_{\mathrm{s}}}
\newcommand{\rhos}{\rho_{\mathrm{s}}}

\newcommand{\sigmam}{\sigma/m}

\renewcommand{\vec}[1]{\bmath{#1}}

\newcommand{\kps}{\, {\rm km}/{\rm s}}

\newcommand{\kpc}{\, {\rm kpc}}

\newcommand{\cmspg}{\, {{\rm cm}^2/{\rm g}}}

\newcommand{\degrees}{^{\circ}}

\newcommand{\Msun}{\,\mathrm{M}_{\odot}}

\newcommand{\pc}{\mathrm{pc}}

\providecommand{\ga}{\gtrsim}
\providecommand{\la}{\lesssim}

\newcommand{\hsi}{h_\mathrm{si}}
\newcommand{\mpar}{m_\mathrm{p}}
\newcommand{\hMpc}{h^{-1} \, {\mathrm{Mpc}}}
\newcommand{\hkpc}{h^{-1} \, {\mathrm{kpc}}}
\newcommand{\hMsun}{h^{-1} \, {\mathrm{M}}_\odot}

\begin{document}

\bibliographystyle{mn2e}

\title [Self-Interacting Dark Matter Simulations I]{Cosmological
Simulations with Self-Interacting Dark Matter I: Constant Density Cores and
Substructure}

\author[Rocha et al.] 
{Miguel Rocha$^1$\thanks{E-mail: rocham@uci.edu},
Annika H. G. Peter$^1$,
James S. Bullock$^1$,
Manoj Kaplinghat$^1$,\newauthor
Shea Garrison-Kimmel$^1$, 
Jose O\~norbe$^1$,
and Leonidas A. Moustakas$^2$\\
$^1$Center for Cosmology, Department of Physics and Astronomy, University of
California, Irvine, CA 92697-4575, USA\\
$^2$Jet Propulsion Laboratory, California Institute of Technology, Pasadena, 
CA 91109, USA\\
}

\date{\today}

\maketitle

\begin{abstract} 
We use cosmological simulations to study the effects of self-interacting dark
matter (SIDM) on the density profiles and substructure counts of dark matter
halos from the scales of spiral galaxies to galaxy clusters, focusing explicitly
on models with cross sections over dark matter particle mass $\sigma/m = 1$ and
$0.1$ $\hbox{cm}^2/\hbox{g}$.  Our simulations rely on a new SIDM N-body
algorithm that is derived self-consistently from the Boltzmann equation and that
reproduces analytic expectations in controlled numerical experiments.   We find
that well-resolved SIDM halos have constant-density cores, with significantly
lower central densities than their CDM counterparts.   In contrast, the subhalo
content of SIDM halos is only modestly reduced compared to CDM,   with the 
suppression greatest 
for large hosts and small halo-centric distances. Moreover, the large-scale
clustering and halo circular velocity functions in SIDM are effectively
identical to CDM, meaning that all of the large-scale successes of CDM are
equally well matched by SIDM.  
 From our largest cross section runs 
we are able to extract scaling relations for core sizes and central densities
over a range of halo sizes and find a strong correlation between the core radius
of an SIDM halo and the NFW scale radius of its CDM counterpart. We construct a
simple analytic model, based on CDM scaling relations, that captures all aspects
of the scaling relations for SIDM halos.  Our results show that halo core
densities in $\sigma/m = 1 \, \hbox{cm}^2/\hbox{g}$ models are too low to match
observations of galaxy clusters, low surface brightness spirals (LSBs), and
dwarf spheroidal galaxies.  However, SIDM with  $\sigma/m \simeq 0.1 \,
\hbox{cm}^2/\hbox{g}$ appears capable of reproducing reported core sizes and
central densities of dwarfs, LSBs, and galaxy clusters without the need for
velocity dependence. Higher resolution simulations over a wider range of masses
will be required to confirm this expectation. We discuss constraints arising
from the Bullet cluster observations, measurements of dark matter density on
small-scales and subhalo survival requirements, and show that SIDM models with
$\sigma/m \simeq 0.1 \, \hbox{cm}^2/\hbox{g} \simeq 0.2\, \hbox{barn}/\hbox{GeV}$ 
are consistent with all observational constraints. 
\end{abstract}

 
\section{Introduction}\label{intro.sec}
There is significant evidence that some form of dark matter dominates the
gravitating mass in the universe and its abundance is known to great precision
\citep{komatsu11}.   The most popular candidate for dark matter is the class of
weakly interacting massive particles (WIMPs), of which supersymmetric
neutralinos are examples \citep{steigman1985,griest1988,jungman1996}.  WIMPs are
stable, with negligible self-interactions, and are non-relativistic at
decoupling (``cold``).  It is important to recognize that of these
characteristics, 
 it is primarily their coldness that is well tested via its association with
significant small-scale power.   Indeed, WIMPs are the canonical Cold Dark
Matter (CDM) candidate.  Cosmological models based on CDM reproduce the spatial
clustering of galaxies on large scales quite well \citep{reidetal10} and even 
 the clustering of galaxies on $ \sim 1$ Mpc scales appears to match that
expected for CDM {\em subhalos} 
\citep{kravtsov2004,conroy2006,trujilloGomezetal11,reddick2012}.

Beyond the fact that the universe appears to behave as expected for CDM on large
scales,  we have few constraints on the microphysical parameters of the dark
matter, especially those that would manifest themselves at the high densities
associated with cores of galaxy halos.     It is worth asking what (if anything)
about vanilla CDM can change without violating observational bounds.  In this
paper we use cosmological simulations to explore the observational consequences
of a CDM particle that is strongly self-interacting, focusing specifically on
the limiting case of velocity-independent, elastic scattering.

Dark matter particles with appreciable self-interactions have been discussed in
the literature for more than two decades
\citep{carlson92,machacek93,delaix95,spergelandsteinhardt00,firmani00}, and are
now recognized as generic consequences of hidden-sector extensions to the
Standard Model \citep{pospelov08, arkanihamed09, ackerman09, feng09, feng10a, loeb11}. 
Importantly, even if
dark sector particles have no couplings to Standard Model particles they
might experience strong interactions with themselves, mediated by dark gauge
bosons
(see \citealt{feng10} and \citealt{peter12} for reviews).   {\em The implication
is that  astrophysical constraints associated with the small-scale clustering of
dark matter may be the only way to test these scenarios}.

Phenomenologically, self-interacting dark matter (SIDM) is attractive because it
offers a means to lower the central densities of galaxies without destroying the
successes of CDM on large scales.
Cosmological simulations that contain only CDM indicate that dark-matter halos
should be cuspy and with (high) concentrations that correlate with the collapse
time of the halo \citep{nfw97, bullock2001, wechsler2002}.  This is inconsistent
with observations of galaxy rotation curves, which show that galaxies are less
concentrated and less cuspy than predicted in CDM simulations
\citep[e.g.][]{floresandprimack94,simonetal05,
kuzioetal08, bloketal10, dutton2011, kuzioetal11,ohetal11a,
walkerandpenarrubia11, saluccietal12, castignanietal12}.  Even for clusters of
galaxies, the density profiles of the host dark-matter halos appear in a number
of cases to be shallower than predicted by CDM-only structure simulations, with
the total (dark matter + baryons) density profile in a closer match to the CDM
prediction for the dark matter alone \citep[e.g.][]{sandetal04, sand2008,
newmanetal09, newmanetal11,coe2012,umetsu2012}.   

One possible answer is feedback.   In principle, the expulsion of gas from
galaxies can result in lower dark matter densities compared to dissipationless
simulations, and thus bring CDM models in line with observations
\citep{governato2010,ohetal11b,pontzen2011,brooketal12,governato2012}.  However,
a new level of concern exists for dwarf spheroidal galaxies \citep{mbketal11a,
ferreroetal11,mbketal11b}.  Systems with $M_* \sim 10^6 \Msun$  appear to be
missing $\sim 5 \times 10^7 \Msun$ of dark matter compared to standard CDM
expectations  \citep{mbketal11b}.  It is difficult to understand how feedback
from such a tiny amount of star formation could have possibly blown out enough
gas to reduce the densities of dwarf spheroidal galaxies to the level required
to match observations
(\citealt{mbketal11b,teyssieretal2012,zolotov2012,penarrubia2012};
Garrison-Kimmel et al., in preparation).  

\cite{spergelandsteinhardt00} were the first to discuss SIDM in the context of
the central density problem (see also \citealt{firmani00}).  
The centers of SIDM halos are expected to have constant density isothermal cores that arise
as kinetic energy is transmitted from the hot outer halo
inward \citep{balberg2002,colinetal02,ahn2005,koda2011}.
This can happen if the cross section over mass of the dark matter particle,
$\sigmam$, is large enough for there to be a relatively high
probability of scattering over a time $t_{\rm age}$ comparable to the age of the
halo:  $\Gamma \, t_{\rm age} \sim 1$, where $\Gamma$ is the scattering rate per particle. 
The rate will vary with local dark matter density $\rho(r)$ as a function of
radius $r$ in a dark halo as
\begin{equation}\label{eq:gamma}
 \Gamma(r) \simeq \rho(r) (\sigma/m) v_{\mathrm{rms}}(r) \, ,
\end{equation}
up to order unity factors, where $v_{\mathrm{rms}}$ is the rms speed of
dark-matter particles.   Based on rough analytic arguments,
\citet{spergelandsteinhardt00} suggested  $\sigmam \sim
0.1-100 \ \cmspg$ would produce observable consequences in the cores of halos. 

Numerical simulations have confirmed the expected phenomenology of core
formation \citep{bukert00} though
   \citet{kochanek&white00} emphasized the possibility that SIDM halos could
eventually
become {\em more} dense than their CDM counterparts as a result of eventual heat
flux from the inside out (much like core collapse globular clusters).  However
this process is suppressed when merging from hierarchical formation is included
\citep[for a discussion see][]{ahn2005}.  We do not see clear signatures of core
collapse in the halos we analyzed for $\sigmam=1 \ \cmspg$. 

The first cosmological simulations aimed at understanding dwarf densities were
performed by \citet{daveetal01} who used a small 
volume ($4 \hMpc$ on a side) in order to focus computational power on dwarf
halos.  They concluded that 
$\sigmam = 0.1-10 \cmspg$ came close to reproducing core densities of
small galaxies, favoring the upper end of that range but
not being able to rule out the lower end due to resolution.  Almost
concurrently, \citet{yoshida00} ran cosmological simulations focusing on
the cluster-mass regime.  Based on the estimated core size of cluster CL
0024+1654, they concluded that
cross sections no larger than $\sim 0.1 \ \cmspg$ were allowed, raising doubts
that constant-cross-section SIDM models could be consistent with observations of
both dwarf galaxies and clusters.  

These concerns were echoed by \citet{miralda2002} who suggested that SIDM halos
would be significantly more spherical than observed for galaxy clusters. 
Similarly, \citet{gnedinandostriker01} argued that SIDM would lead to excessive
sub halo evaporation in galaxy clusters.  More recently, the merging cluster
system known as the Bullet Cluster has been used to derive the limits (68\% C.L.) 
$\sigmam < 0.7  \cmspg$ \citep{randalletal08} based on evaporation of 
dark matter from the subcluster and $\sigmam <1.25  \cmspg$  \citep{randalletal08} based on 
the observed lack of offset between the bullet subcluster mass peak and the galaxy light centroid.  
In order to relax this apparent tension between what was required to match dwarf densities and 
the observed properties of galaxy clusters, velocity dependent cross sections that 
diminish the effects of self-interaction in cluster environments have been
considered \citep{firmani00,colinetal02,feng09,loeb11,vogelsberger12}.

There are a few new developments that motivate us to revisit constant SIDM cross
sections on the order of $\sigmam \sim 1 \cmspg$.    For example, the cluster
(CL 0024+1654) used by \citet{yoshida00} to place one of the tightest limits at
$\sigmam = 0.1$,  is now recognized as an ongoing merger along the line of sight
\citep{czoske2001,czoske2002,zhang2005,jee2007,jee2010,umetsu2010}.  This calls
into question its usefulness as a comparison case for non-merging cluster
simulations.  In a companion paper (Peter, Rocha, Bullock and Kaplinghat, 2012)
we use the same simulations described here to show that published constraints on
SIDM based on halo shape comparisons are significantly weaker than previously
believed. 
  Further, the results presented below clearly demonstrate that the tendency for
subhalos to evaporate
in SIDM models \citep{gnedinandostriker01} is not significant for $\sigmam \sim
1 \cmspg$.
Finally (and related to the previous point), the best numerical analysis  of the Bullet Cluster
\citep{randalletal08} used initial cluster density profiles that were unmotivated cosmologically with 
central densities about a factor of two too high for the SIDM cross sections considered
(producing a scattering rate that is inconsistently high).  Based on this observation, the bullet 
cluster constraint based on evaporation of dark matter from the subcluster should be relaxed 
since the amount of subcluster mass that becomes unbound is directly proportional to the 
density of dark matter encountered in its orbit. 
Moreover, their model galaxies were placed in the cluster halo potentials
without subhalos surrounding them, an assumption (based on analytic estimates
for SIDM subhalo evaporation) that is not supported by our simulations. This could 
affect the constraints based on the (lack of) offset between dynamical mass and light. 
Thus we believe that the bullet cluster constraints as discussed above are likely only 
relevant for models with $\sigmam > 1 \cmspg$.  However, the constraints could be 
made significantly stronger by comparing SIDM predictions to the densities inferred 
from the convergence maps since the central halo densities for 
$\sigmam \simeq 1 \cmspg$ are significantly lower than the CDM predictions, as we
show later. 

Given these motivations, we perform a set of cosmological simulations with both
CDM and SIDM.  For SIDM we ran
$\sigmam = 1$ and $0.1 \cmspg$ models (hereafter SIDM$_1$ and SIDM$_{0.1}$), {\em i.e.}, models
that we have argued pass the Bullet cluster tests.  
Our simulations provide us with a
sample of halos that span a mass range much larger than either
\citet{daveetal01} or \citet{yoshida00} both with and without
self-interactions. 

One of the key findings from our simulations is that the core sizes are expected
to scale approximately as a fixed fraction of the NFW scale radius the halo
would have in the absence of scatterings. We can see where this scaling arises
from a quick look at Equation~\ref{eq:gamma}. This equation allows us to argue
that the radius ($r_1$) below which we expect dark matter particles (on average)
to have scattered once or more is set by: 
\begin{equation}\label{eq:onescatter}
 \rhos f(r/\rs) v_\mathrm{rms} \propto \frac{\Vmax^2}{\Rmax^2} f(r_1/\rs) \Vmax
= \rm constant \enspace,
\end{equation}
where $f(x)$ is the functional form of the NFW (or a related) density profile. In
writing the above equation we have assumed that the density profile for SIDM is
not significantly different from CDM at $r_1$, something that we verify through
our simulations. Now, since CDM enforces a $\Vmax-\Rmax$ relation such that
$\Vmax\propto \Rmax^{1.4-1.5}$, we see that the solution to $r_1/\rs$ is going
to be only mildly dependent on the halo properties. We develop an analytic model
based on this insight later, but this is the underlying reason for why we find
core sizes to be a fixed fraction of the NFW scale radius of the same halo in
the absence of scatterings.

The major conclusion we reach based on the simulations and the analytic 
model presented here is that a self-interacting dark matter model with a cross-section over 
dark matter particle mass $\sim 0.1 \cmspg$ would be capable of reproducing 
the core sizes and central densities observed in dark matter halos {\em at all 
scales}, from clusters to dwarf spheroidals, without the need for 
velocity-dependence in the cross-section.

In the next section, we discuss our new algorithm to compute the
self-interaction probability for N-body particles, derived self-consistently
from the Boltzmann equation. We discuss this new algorithm in detail in Appendix 
\ref{appendixA}. In \S \ref{implementation.sec}, we show how this algorithm is
implemented in the publicly available code GADGET-2
\citep{springel05}. We run tests that show that our algorithm gets the correct
interaction rate
and post-scattering kinematics. The results of these tests are in \S
\ref{test.sec}. The cosmological simulations with this new algorithm are
described in detail in
\S \ref{sims.sec}. In \S \ref{prelim.sec} we provide some preliminary
illustrations of our simulation snapshots and in \ref{lss:sec} we demonstrate
that the large-scale statistical 
properties of SIDM are identical to CDM.   In \S
\ref{halos.sec} we present the properties of individual SIDM$_1$ and
SIDM$_{0.1}$ halos and compare them to the their CDM counterparts. 
In \S \ref{subhalos.sec} we discuss the subhalo mass functions 
in our SIDM and CDM simulations and show that SIDM$_1$ subhalo mass functions are 
very close to that of CDM in the range of halo masses we can resolve. 
We provide scaling relations for the 
SIDM$_1$ halo properties in \S \ref{scaling.sec} and in \S \ref{analytic.sec} we 
present an analytic model that reproduces these scaling relations as well as 
the absolute densities and core radii of 
SIDM$_1$ halos. We use these scaling relations and the analytic model to make 
a broad-brush comparison to observed data in \S \ref{discussion.sec}. We present a
summary together with our final conclusions in \S \ref{sumandconc.sec}.

\section{Simulating Dark Matter Self Interactions}
\label{implementation.sec}

Our simulations rely on  a new algorithm for modeling self-interacting dark
matter with N-body simulations.  Here we introduce our approach and provide a
brief summary.  In Appendix \ref{appendixA} we derive the algorithm explicitly
starting with the Botlzmann equation and give details for general
implementation.

In N-body simulations, the simulated (macro)particles represent an ensemble of
many dark-matter particles.  Each simulation particle of mass $\mpar$ can be
thought of as a patch of dark-matter phase-space density. In our treatment of
dark matter self-scattering, the phase space patch of each particle is
represented by a delta function in velocity and a spatially extended kernel
$W(r,\hsi)$, smoothing out the phase space in configuration space on a
self-interaction smoothing length $\hsi$.  The value of $\hsi$ needs to be
set by considering the physical conditions of the problem (see \S
\ref{test.sec}) as it specifies the range over which N-body particles can affect
each other via self-interactions.  In principle, $\hsi$ could be different for
each particle and vary depending on the local density, but in the simulations
presented here we fix $\hsi$ to be the same for all particles in a given
simulation, setting the size of $\hsi$ according the lowest densities at which
self-interactions are effective for a given cross section.

When two phase-space patches overlap, we need to calculate the pairwise
interaction rate between them.  We do so by considering the ``scattering out''
part of the Boltzmann collision term in Equation (\ref{eq:boltzmann}) and Eqs.
(\ref{eq:pair1})-(\ref{eq:pair6}).  The implied rate of scattering of an
N-body particle $j$ off of a target particle $i$ of mass $\mpar$ is
\begin{equation}
\label{gameq.eq}
\Gamma(i|j)  =  (\sigma/m) \mpar | \mathbf{v}_i - \mathbf{v}_j | g_{ji} \, ,
\end{equation}
where $g_{ji}$ is a number density factor that accounts for the overlap of the
two particles' smoothing kernels:~\footnote{This equation applies only if $\hsi$
is the same for both particles.  See Appendix A for the general form.}
\begin{equation}
 g_{ji} = \int_0^{\hsi} d^3 \mathbf{x}^\prime
W(|\mathbf{x}^\prime|,\hsi) W(| \delta \mathbf{x}_{ji} +
\mathbf{x}^\prime|,\hsi) \, .
\end{equation}
The probability that such an interaction occurs in a time step $\delta t$ is 
\begin{equation}
\label{probeq.eq}
P(i|j) = \Gamma(i|j) \, \delta t \, ,
\end{equation}
and the total probability of interaction between N-body particles $i$ and $j$ is
\begin{equation}
\label{totalProb.eq}
P_{ij} = \frac{P(i|j) + P(j|i)}{2}.
\end{equation}
Specifically, $P_{ij}$ is the probability for
a macroparticle representing a patch of phase space around
$(\mathbf{x}_j,\mathbf{v}_j)$ to interact with a target particle representing a
patch of phase space around $(\mathbf{x}_i,\mathbf{v}_i)$ in a time $\delta t$. 

We determine if particles interact by drawing a random number
for each pair of particles that are close enough for the probability of
interaction to be greater than zero.
If a pair does scatter, we do a Monte Carlo for the new velocity
directions, populating these parts of the phase-space and deleting
the two particles at their initial phase-space locations. Note that by virtue of
populating the new phase space regions, we are taking care of the ``scattering
in'' term of the collision integral in Equation (\ref{eq:boltzmann}). We avoid
double counting by only accounting for $P_{ij} = P_{ji}$ once during a given
time-step $\delta t$.  
In the limit of a large number of macroparticles, the total interaction
probability for each particle $i$ should approach 
\begin{equation}
\label{probTotal.eq}
P_i = \sum_{j} P_{ij} \,.
\end{equation}
We show in \S 3 that this approach correctly reproduces the expected number of
scatterings in a idealized test case. 

Our method for simulating scattering differs from previous approaches in a few
key ways.  It is most similar to that of \citet{daveetal01}
in that we both directly consider interactions between pairs of phase-space
patches and rely on a scattering rate similar in form to Equation
\ref{gameq.eq}.   The difference is that their geometric factor $g_{ji}$ is not
the same---our factor arises explicitly from the overlap in patches of phase
space between neighboring macroparticles, as derived from the collision term in
the Boltzmann equation (see Appendix A for details). Other authors determine the
scattering rate $\Gamma$ of individual phase-space patches based on estimates of
the local mass density (typically using some number of nearest neighbors or
using an SPH kernel).
The Monte Carlo is then based on an estimated scattering rate of an individual
particle on the background, and a scattering partner is only chosen after a
scattering event is determined to have occurred  \citep{kochanek2000, yoshida00,
colinetal02,randalletal08}.  The  scattering probability in this latter approach
is not symmetric.  For macroparticles of identical mass, $P(i|j) = P(j|i)$
explicitly in our approach, but not the other approach because the density
estimated at the position of macroparticle $i$ need not be the same as that
estimated at the position of particle $j$.  In the future, there should be a
direct comparison among these scattering algorithms to determine if they yield
consistent results.

\begin{figure} 
\begin {center}
\includegraphics[width=0.5\textwidth]{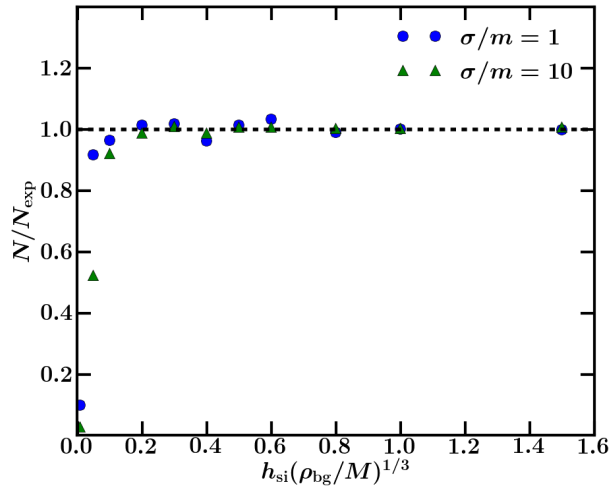}
\end {center} 
\caption{Fraction of the expected total number of interactions that are computed
in our test simulation as a function of the self-interaction
smoothing length. The self-interaction cross section for each run is shown in
units of cm$^2$/g in the legend.  The code converges to the expected number of
interactions when the smoothing length approaches the background inter-particle
separation, i.e. when $h_\mathrm{si} (\rho_\mathrm{bg}/\mpar)^{1/3} \gtrsim
0.2$.}
\label{hconvtestFig} 
\end{figure}

We have implemented our algorithm in the publicly available version of the
cosmological simulation code GADGET-2 \citep{springel05}. GADGET-2 computes the
short-range gravitational interactions by means of a hierarchical multipole
expansion, also known as a tree algorithm. Particles are grouped hierarchically
by a repeated subdivision of space, so their gravitational contribution can be
accounted by means of a single multipole force computation. A cubical root node
encompasses the full mass distribution.  The node is repeatedly subdivided into
eight daughter nodes of half the side length each (an oct-tree) until one ends
up with ``leaf'' nodes containing single particles. Forces for a given particle
are then obtained by ``walking'' the tree, opening nodes that are too close for
their multipole expansion to be a correct approximation to their gravitational
contribution. In GADGET-2, spurious strong close encounters by particles are
avoided by convolving the single point particle density distribution with a
normalized spline kernel (``gravitational softening'').  

To implement our algorithm, we take advantage of the tree-walk already build in
GADGET-2, computing self interactions during the calculation of the
gravitational interactions.  For this to work we
have to modify the opening criterion such that nodes are opened if they are able
to have particles closer than $2 \hsi$ from a target scatterer (or $h_i + h_j$
if particles have different self-interaction smoothing lengths).  
When computing the probability of interaction we use the same spline kernel used
in GADGET-2 \citep{monaghan&lattazio85}, defined as
\begin{equation}
W(r,h)=\frac{8}{\pi h^3} \left\{
\begin{array}{ll}
1-6\left(\frac{r}{h}\right)^2 + 6\left(\frac{r}{h}\right)^3, &
0\le\frac{r}{h}\le\frac{1}{2} ,\\
2\left(1-\frac{r}{h}\right)^3, & \frac{1}{2}<\frac{r}{h}\le 1 ,\\
0 , & \frac{r}{h}>1 .
\end{array}
\right. \label{eqkernel}
\end{equation}

If a pair interacts we give both particles a kick consistent with an elastic
scattering that is isotropic in the center of mass frame.  The post-scatter
particle velocities are 
\begin{align}
\mathbf{v}_0^\prime &= \mathbf{v}_c + \frac{m_1}{m_0+m_1}V \mathbf{e},
\nonumber \\
\mathbf{v}_1^\prime &= \mathbf{v}_c - \frac{m_0}{m_0+m_1}V \mathbf{e},
\end{align}
where $\mathbf{v}_c$ is the center of mass velocity, $V$ is the relative speed
of the particles (conserved for elastic collisions) and $\mathbf{e}$ is a
randomly chosen direction.

The time-step criterion is also modified to assure that the scattering
probability for any pair of particles is small, $P = \Gamma \ \delta t
<< 1$.  An individual particle time-step is decreased by
a factor of 2 if during the last tree-walk the maximum probability of
interaction for any pair involving such a particle was $P_\mathrm{max} >
0.2$. Once a particle time-step is modified due to the previous restriction,
if $P_\mathrm{max} < 0.1$ for such a particle and its current time-step is
smaller than the one
given by the standard criterion on GADGET-2, we increase it by a factor of 2.

\begin{table*}
\label{sims.tab}
{\bf Table 1:} Simulations discussed in this paper.\\
\centering
\begin{tabular}{lcccccc}
Name & Volume & Number of Particles & Particle Mass &Force Softening &
Smoothing Length & Cross-section \\
& $L_\mathrm{Box}$  [$\hMpc$] & $N_\mathrm{p}$ & $m_\mathrm{p}$  [$\hMsun$] &
$\epsilon$  [$\hkpc$] & $h_\mathrm{si}$  [$\hkpc$] & $\sigmam$ 
[$\cmspg$] \\
\hline \hline
CDM-50 & $50$ & $512^3$ & $6.88\times10^7$ & $1.0$ & $-$ & 0\\
CDM-25 & $25$ & $512^3$ & $8.59\times10^6$ & $0.4$ & $-$ & 0\\ 
CDM-Z11 & $(3 R_{\rm vir})$* & $2.5\times10^6$* & $1.07\times10^6$*& $0.3$ & $-$
& 0\\
CDM-Z12 & $(3 R_{\rm vir})$* & $5.6\times10^7$* & $1.34\times10^5$* & $0.1$ &
$-$ & 0\\ 
\hline
SIDM$_{0.1}$-50 & $50$ & $512^3$ & $6.88\times10^7$ & $1.0$ & $2.8 \ \epsilon$ &
0.1\\
SIDM$_{0.1}$-25 & $25$ & $512^3$ & $8.59\times10^6$ & $0.4$ & $2.8 \ \epsilon$ &
0.1\\
SIDM$_{0.1}$-Z11 & $(3 R_{\rm vir})$* & $2.5\times10^6$* & $1.07\times10^6$* &
$0.3$ & $2.8 \ \epsilon$ & 0.1\\
 SIDM$_{0.1}$-Z12 & $(3 R_{\rm vir})$* & $5.6\times10^7$* & $1.34\times10^5$* &
$0.1$ & $1.4 \ \epsilon$ & 0.1 \\
 \hline
SIDM$_1$-50 & $50$ & $512^3$ & $6.88\times10^7$ & $1.0$ & $2.8 \ \epsilon$ & 1\\
SIDM$_1$-25 & $25$ & $512^3$ & $8.59\times10^6$ & $0.4$ & $2.8 \ \epsilon$ & 1\\
SIDM$_1$-Z11 & $(3 R_{\rm vir})$* & $2.5\times10^6$* & $1.07\times10^6$* & $0.3$
& $2.8 \ \epsilon$ & 1\\
 SIDM$_1$-Z12 & $(3 R_{\rm vir})$* & $5.6\times10^7$* & $1.34\times10^5$* &
$0.1$ & $1.4 \ \epsilon$ &
1\\
\end{tabular}
\vskip 0.5 cm *Note: The Z11 and Z12 runs are zoom simulations with
multiple particle species concentrating on halos of mass $M_{\rm vir} = 5 \times
10^{11}$ M$_\odot$ and $1.0 \times 10^{12}$ M$_\odot$, respectively (no $h$).  
The volumes listed refer to the number of virial radii used to find the
Lagrangian volumes associated with the zoom.  The particle properties listed 
are for the
highest resolution particles only.  
\end{table*}

\section{Test of the SIDM Implementation}
\label{test.sec}

Before performing cosmological simulations, we carried out a controlled test of
the implementation in order to make sure the scattering rate and kinematics are
correctly followed in the code, and to determine the optimum value of the SIDM
softening kernel length $\hsi$.  The simplest and cleanest scenario for testing
our
implementation consists of a uniform sphere of particles moving through a
uniform field of stationary background particles. The coordinate system is
defined such as the sphere is moving along the positive z-direction with
constant velocity $v_s$. The particles forming the sphere and the particles
forming the background field are tagged as different types within the code and
here we will refer to them simply as \textit{sphere} (s) and \textit{background}
(bg) particles
respectively. We only allow scatterings involving two different types of
particles (i.e. sphere-background interactions only) and turn off gravitational
forces among all of the particles. Furthermore all particles have the same mass
$\mpar$.


The expected number of interactions for this case is given by

\begin{equation}
\label{expNinteractEq} N_{exp}(t) = \sum_{i\in \textrm{s}, j \in \textrm{bg}}
P_{ij}= N_s (\sigma/m) \rho_\mathrm{bg} v_s \ t \,
\end{equation}
where $N_s$ is the total number of Sphere particles, $\rho_\mathrm{bg}$ is
the density of the background field and $t$ is the elapsed time from the
begining of the simulation. From this experiment we have found  that the number
of interactions computed by the code depends on the self-interaction smoothing
length $h_\mathrm{si}$ (see Figure \ref{hconvtestFig}), which is fixed to be 
the same for all particles in this test. The number of interactions converges to
the expected value given by Equation (\ref{expNinteractEq}) as $h_\mathrm{si}$
becomes comparable to the background inter-particle separation, specifically
when
$h_\mathrm{si} (\rho_\mathrm{bg}/\mpar)^{1/3} \gtrsim 0.2$. For $h_\mathrm{si}
(\rho_\mathrm{bg}/\mpar)^{1/3} \gtrsim 0.5$ the accuracy of the algorithm does
not
improve by much and the time of the calculations increases rapidly, $\propto
h_\mathrm{si}^3$. Apart from the expense, using larger values of $h_\mathrm{si}$
would lead to increasingly non-local interactions among particles, which is
inconsistent with the model under consideration.  

We also check the kinematics of the scatters in this test simulation and
describe the results in Appendix \ref{appendixB}. The resulting kinematics and
number of interactions from our test
simulation agrees well with the expectations from the theory as long as
$h_\mathrm{si} (\rho_\mathrm{bg}/\mpar)^{1/3} \gtrsim 0.2$.   

\begin{figure*}
\begin{center} 
\includegraphics[width=0.45\textwidth]{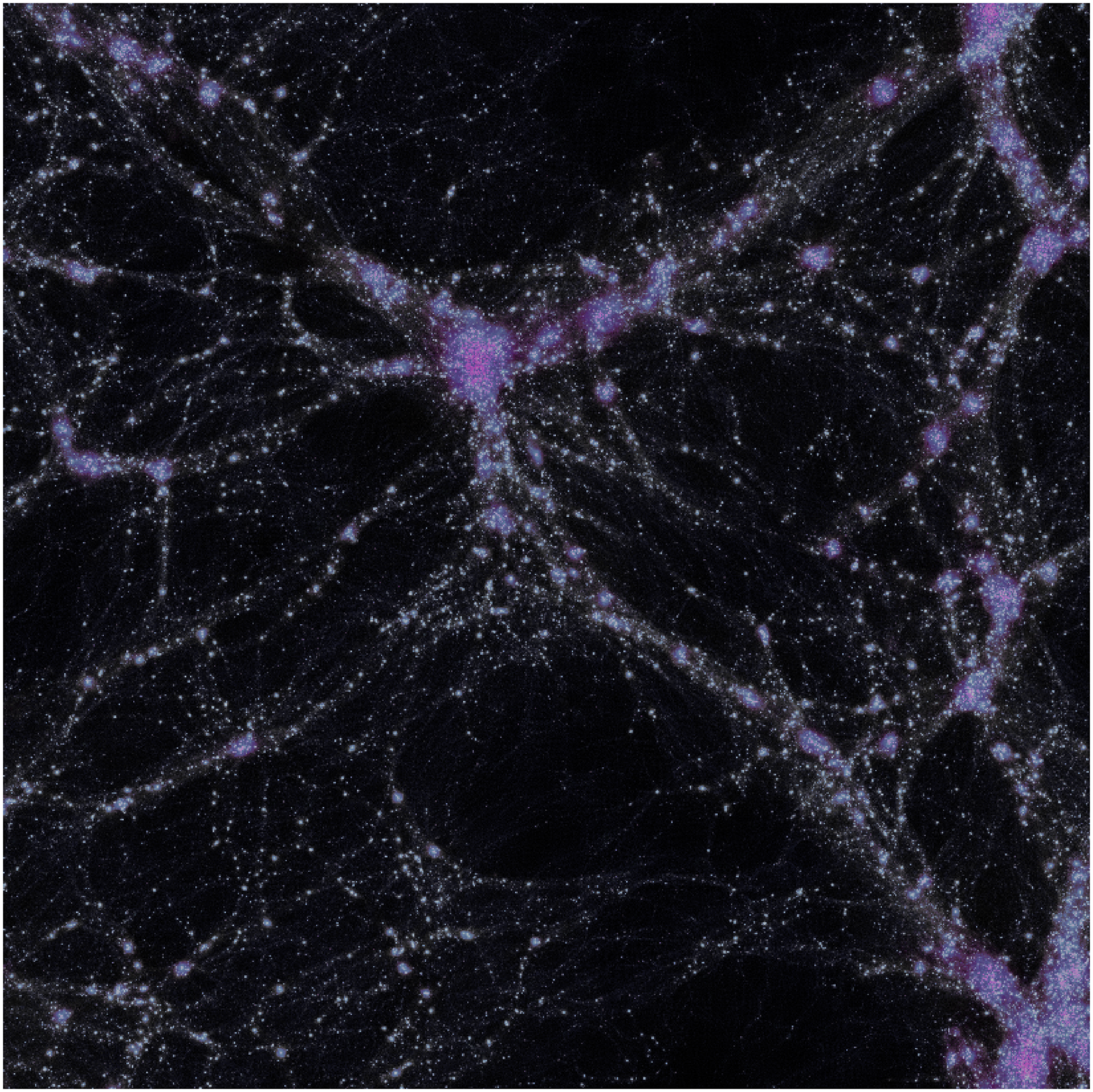}
\includegraphics[width=0.45\textwidth]{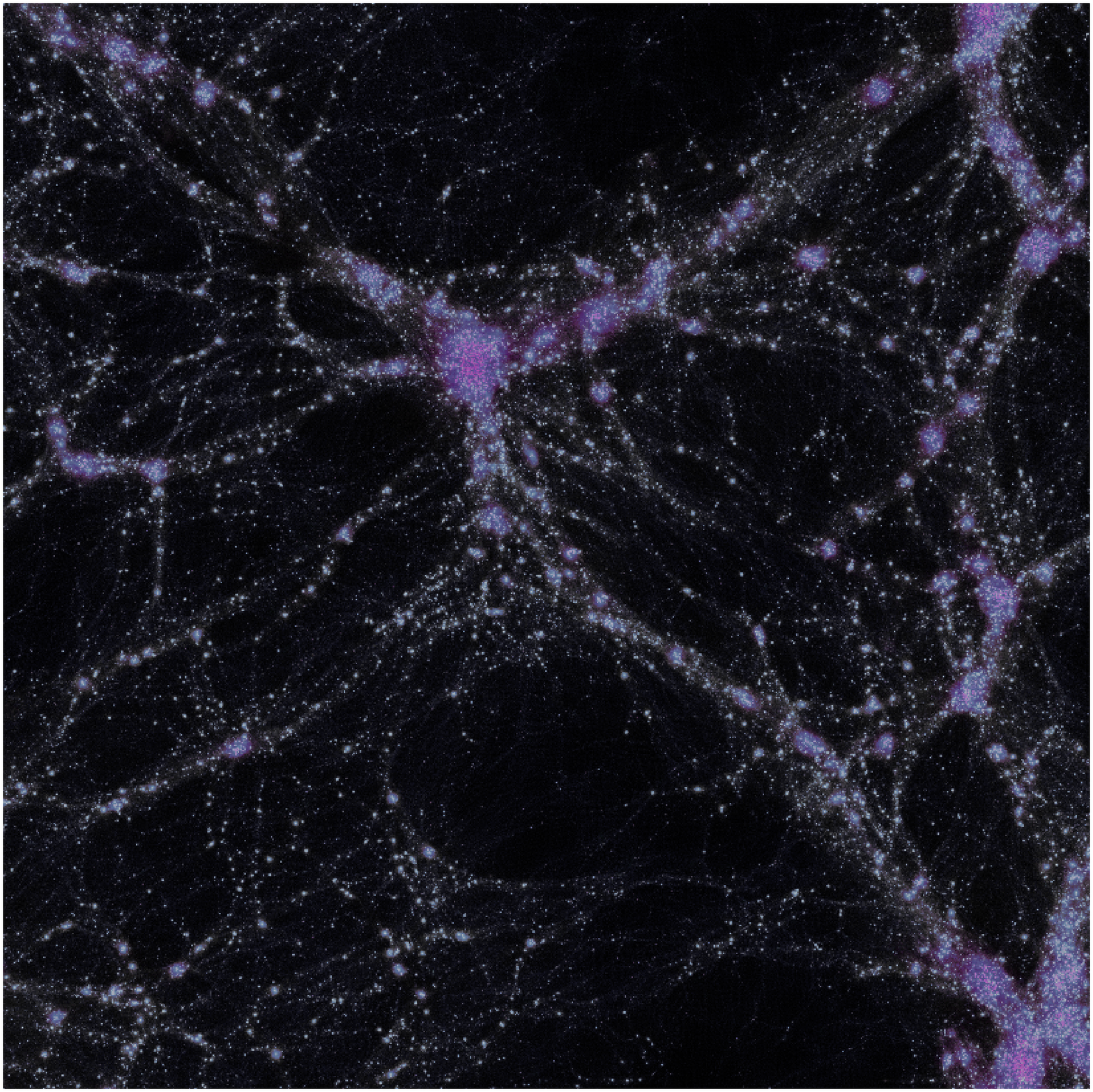}
\vskip.05cm
\includegraphics[width=0.45\textwidth]{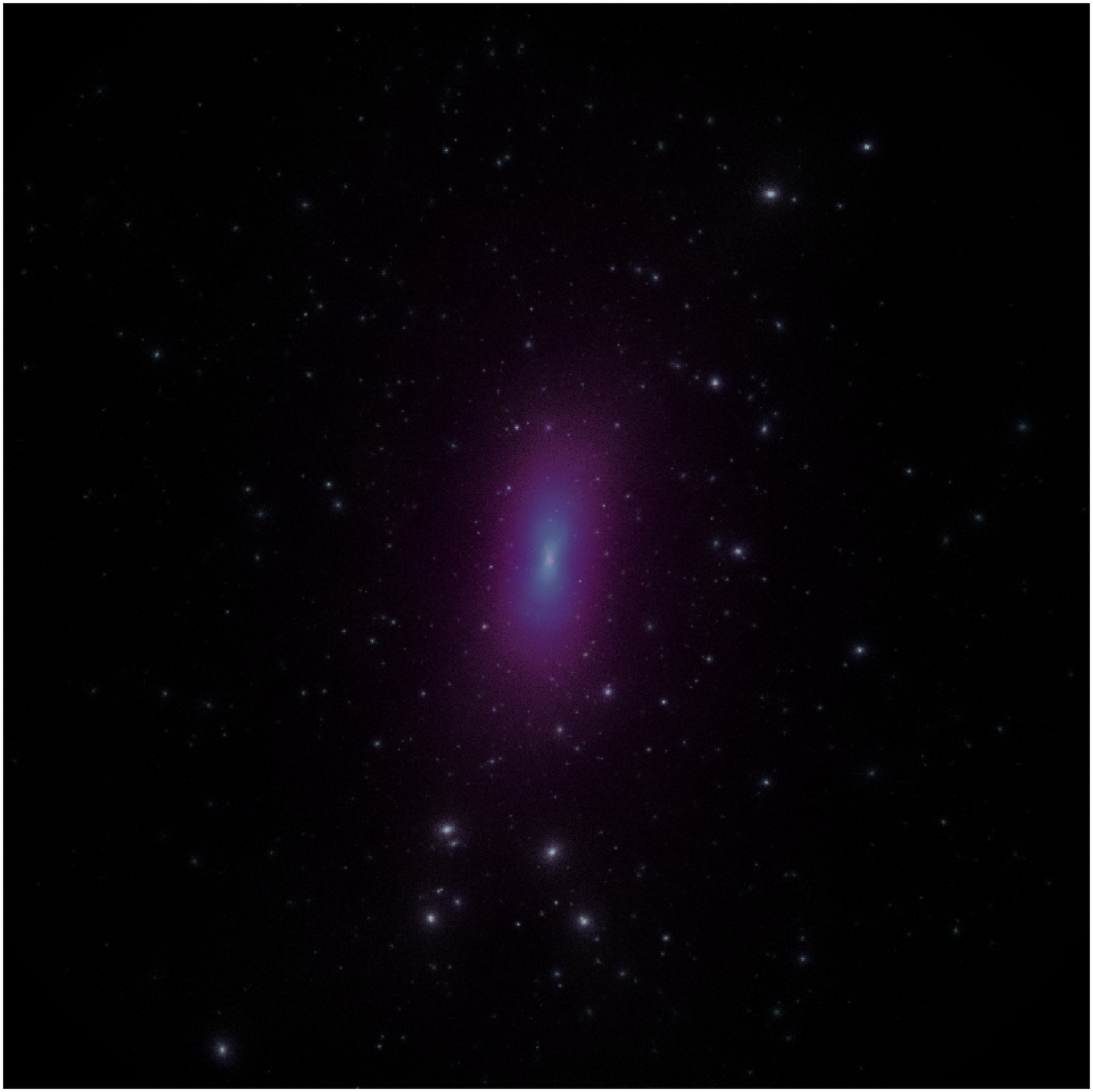}
\includegraphics[width=0.45\textwidth]{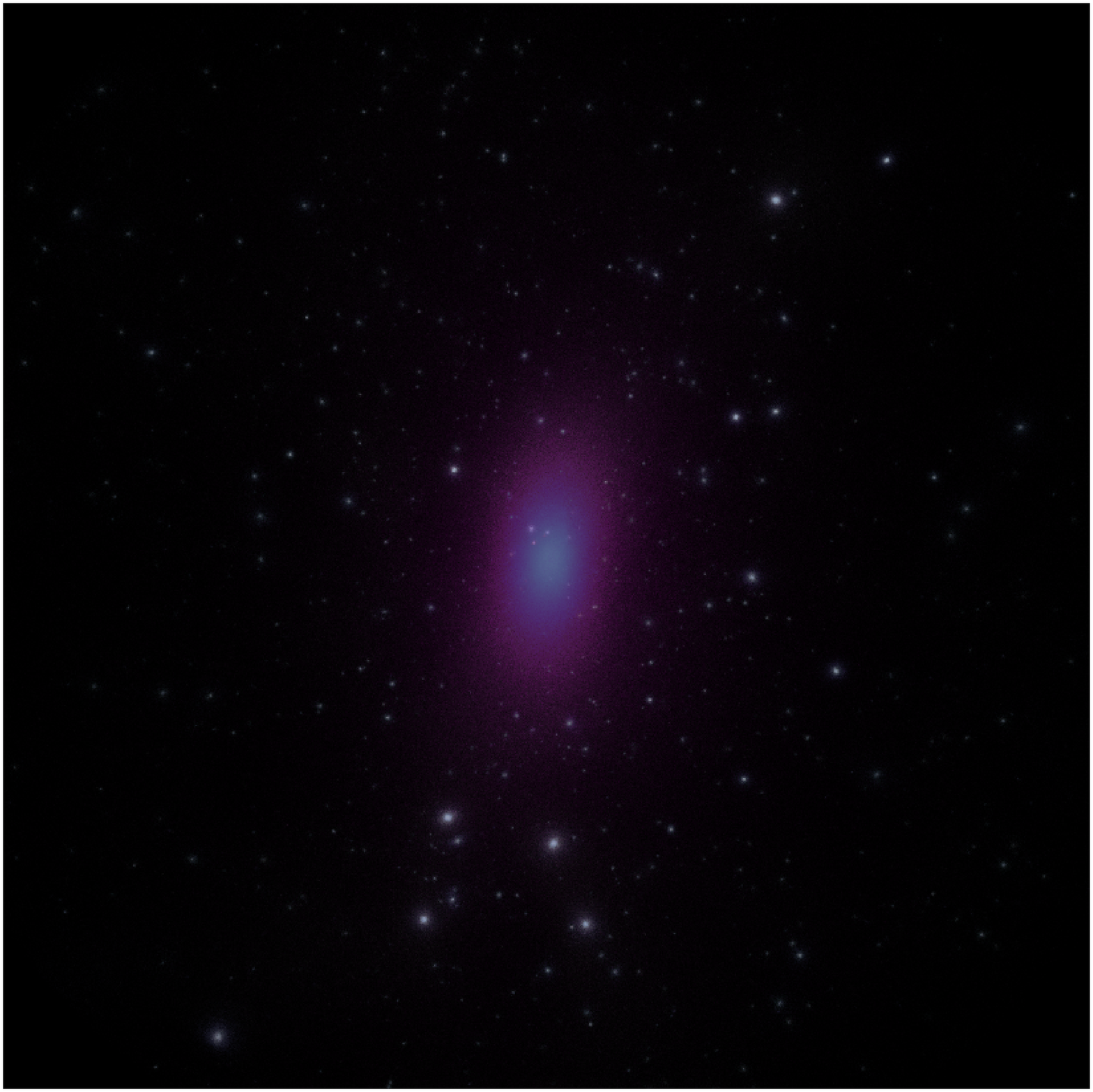}
\end {center} 
\caption{Top: Large scale structure in CDM (left) and SIDM$_1$ (right) shown as
a $50\times50 \, \hMpc$ slice with $10 \, \hMpc$ thickness through our
cosmological simulations. Particles are colored  according to their local
phase-space density. There are no visible differences
between the two cases. Bottom: Small scale structure in a Milky Way mass halo
(Z12) simulated with CDM (left) and SIDM$_1$ (right), including all particles
within $200 \hkpc$ of the halo centers. The magnitude of
the central phase-space density is lower in SIDM because the physical density is
lower {\em and} the velocity dispersion is higher.  The core of the SIDM halo is
also slightly rounder.   Note that substructure content is quite similar except
in the central regions}
\label{Viz.fig} 
\end{figure*}

\begin{figure*}
\begin{center} 
\includegraphics[width=0.498\textwidth]{{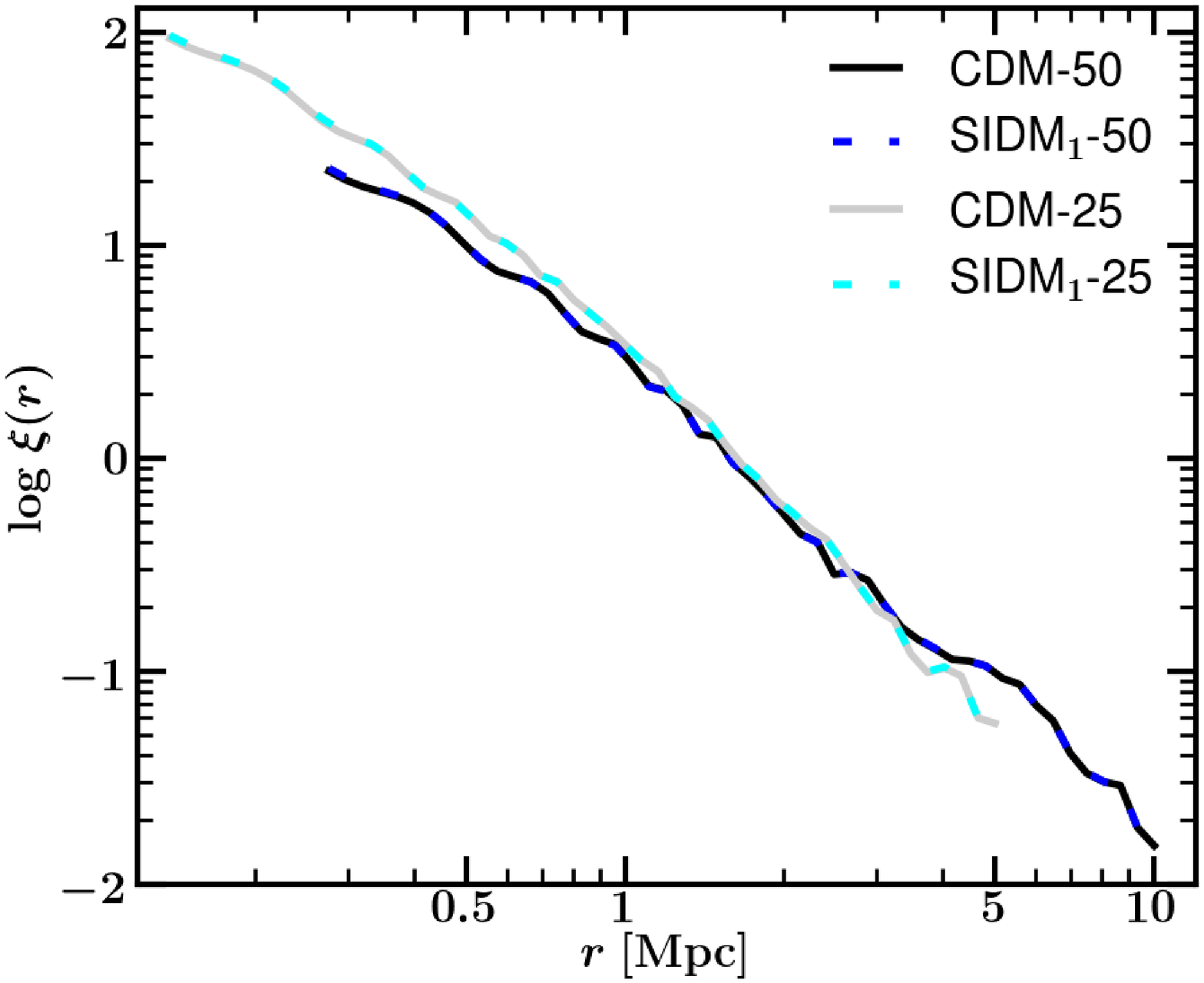}}
\includegraphics[width=0.498\textwidth]{{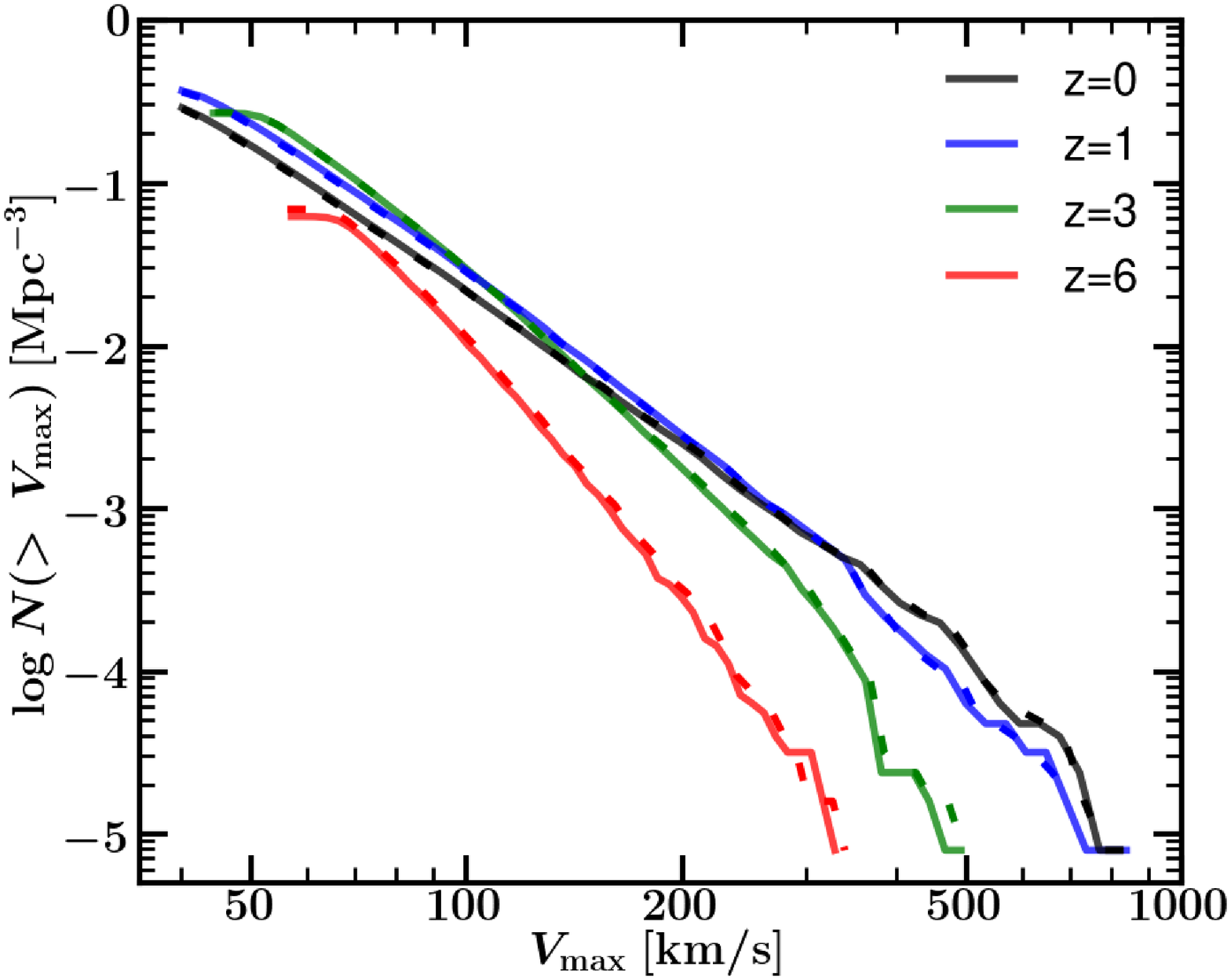}}
\caption{Large-scale characteristics {\em Left:} Dark matter two-point
correlation functions
from our CDM-50 (CDM-25) and SIDM$_1$-50 (SIDM$_1$-25) simulations in
black (grey) and blue (cyan) colors respectively. There are no noticeable
difference between the CDM and SIDM$_1$ dark matter clustering over the scales
plotted.  {\em Right:}
Cumulative number density of dark matter halos as a function of
their maximum circular velocity ($\Vmax$) at different redshifts for our CDM-50
(solid) and SIDM$_1$-50 (dashed) simulations. There are no significant
differences in the $\Vmax$ functions of CDM and SIDM$_1$
at any redshift. }
\label{lss.fig} 
\end{center}
\end{figure*}

\section{Overview of Cosmological Simulations}
\label{sims.sec}

We initialize our cosmological simulations using the Multi-Scale Initial
Conditions (MUSIC) code of
\cite{hahn&abel11}. We have a total of four initial condition sets, each run
with both CDM and SIDM.  The first two are cubic volumes of $25 \hMpc$ and $50
\hMpc$ on a side, each with $512^3$ particles.   As discussed below, these
simulations allow us to resolve the structure of a statistical sample of group
($\sim 10^{13} \Msun$) and cluster ($\sim 10^{14} \Msun$) halos.    
 
The second two initial conditions concentrate computational power on zoom
regions \citep{katz&white93} drawn from
the $50  \hMpc$ box, specifically aimed at exploring the density structure of
two smaller halos, one with virial mass \footnote{We define $\Mvir$ as $\Mvir =
\frac{4}{3} \pi \rho_b \Delta_\mathrm{vir}(z) \Rvir^3$, and $\Rvir$ as
$\tilde{\rho}(\Rvir) = \Delta_\mathrm{vir}(z) \rho_b$. Where
$\tilde{\rho}(\Rvir)$ denotes the
overdensity within $\Rvir$, $\rho_b$ is the background density and
$\Delta_\mathrm{vir}$ the virial overdensity.}
$\Mvir = 7.1 \times 10^{11} \hMsun = 1 \times 10^{12} \Msun$ (Z12) and one with
$\Mvir = 3.5 \times 10^{11} \hMsun = 5 \times 10^{11} \Msun$ (Z11).  
The Z12 run in particular is fairly high resolution, with more than five million
particles in the virial radius.  Table 1 summarizes the simulation parameters.
The cosmology used is based on WMAP7 results
for a $\Lambda$CDM Universe: $h = 0.71$, $\Omega_\mathrm{m} = 0.266$,
$\Omega_{\Lambda} = 0.734$, $\Omega_\mathrm{b} = 0.0449$, $n_\mathrm{s} =
0.963$, $\sigma_8 = 0.801$ \citep{komatsu11}.

Each of our four initial conditions has been evolved from redshift $z = 250$  to
redshift $z = 0$ with collisionless dark matter (labeled CDM) and with
two types of self-interacting dark matter: one with  $\sigmam = 1 \  \cmspg$
(labeled SIDM$_1$) and another with $\sigmam = 0.1 \  \cmspg$
(labeled SIDM$_{0.1}$). We can use the same initial conditions for CDM and SIDM
because at high redshift the low densities and low relative velocities of the
dark matter make self-interactions insignificant.
 Table
\ref{sims.tab} list all the simulations used for this study and detail their
force, mass, and self-interaction resolution.  
In addition to the simulations listed in the table, we also ran the
cosmological boxes with SIDM cross sections $\sigma/m = 0.03\hbox{
cm}^2/\hbox{g}$. We do not present results from these low cross section runs
here because no core density differences were resolved within the numerical
convergence radii of our simulations.


As shown in \S \ref{test.sec} the self-interaction smoothing
length $h_\mathrm{si}$ must be larger than ~20\% the inter-particle separation
in order to
achieve convergence on the interaction rate.   All the work for this
paper was done with a fixed  $h_\mathrm{si}$ for all particles
carefully chosen for each simulation so that the self-interactions are well
resolved at densities a few times to an order of magnitude lower than the lowest
densities for which self-interactions are significant. We have run the
cosmological boxes with different choices for $h_\mathrm{si}$ (changes by
factors of 2 to 4) and have found that our results are unaffected.  We have also
run tests on isolated halos with varying smoothing lengths and again find that
the effects of self-interactions are robust to reasonable changes in
$h_\mathrm{si}$.  

All of our halo catalogs and density profiles are derived using the publicly
available code Amiga Halo Finder (AHF) \citep{knollmannandknebe09}.

\section{Simulation Results}

\subsection{Preliminary Illustrations}
\label{prelim.sec}

Before presenting any quantitative comparisons between our CDM and SIDM runs, we
provide some simulation renderings in order to help communicate the qualitative
differences.

The upper panels of Figure \ref{Viz.fig} show a large-scale comparison: two
($50\times50\times10$)  $\hMpc$ slices from the CDM-50 and SIDM$_1$-50
boxes side-by-side at $z=0$. The structures are color-coded by local phase-space
density ($\propto \rho/v_{\rm rms}^3$). 
It is evident that there are no observable differences in the large-scale
characteristics of CDM and SIDM$_1$.  We discuss this result in more
quantitative terms in \S \ref{lss:sec} but of course this is expected.  The SIDM
models we explore do not have appreciable rates of interaction for densities
outside the cores of dark matter halos.   The upper panels of Figure
\ref{Viz.fig} provide a visual reminder that the SIDM models we consider are
effectively identical to CDM on larges scales.

The differences between CDM and SIDM become apparent only when one considers the
internal structure of individual halos.
The lower panels of Figure \ref{Viz.fig} provide side-by-side images of a
Milky-Way mass halo (Z12) simulated with CDM (left) and 
SIDM$_1$ (right).  SIDM tends to make the cores of halos less dense {\em and}
kinetically hotter (see \S \ref{halos.sec}) and these two differences are
enhanced multiplicatively in the phase-space density
renderings. The central regions of the host halo are also slightly rounder in
the SIDM case (Peter et al. 2012).
Importantly, the difference in substructure characteristics are minimal,
especially at larger radii.  We return to a quantitative description of
substructure differences in \S \ref{subhalos.sec}.

\subsection{Large Scale Structure and Halo Abundances}
\label{lss:sec}

Figure \ref{lss.fig} provides a quantitative comparison of both the clustering
properties (left) and halo abundance evolution (right) between 
our full-box CDM and SIDM$_1$ simulations.  The left panel shows the two-point
function of dark matter particles in both cosmological runs for 
CDM and SIDM$_1$.  There are no discernible differences between SIDM and CDM
over the scales plotted, though of course the different box sizes (and
associated resolutions) mean that the boxes themselves only overlap for a
limited range of scales.  For a given set of initial conditions, however, SIDM
and CDM give identical results.

The right panel of Figure \ref{lss.fig} shows the cumulative number density of
dark-matter halos (including subhalos) as a function of their peak circular
velocity ($\Vmax$) for the CDM-50 (solid) and SIDM$_1$-50 (dashed) simulations
at various redshifts.  Remarkably, this comparison shows no significant
difference either -- indicating that SIDM with cross sections as large as $1
\cmspg$  does not strongly affect the maximum circular velocities of individual
halos.   The two panels of Figure \ref{lss.fig} demonstrate that for large-scale
comparisons, including analyses involving field halo mass functions, SIDM and
CDM yield identical results.  The implication is that observations of
large-scale structure are just as much a ``verification" of SIDM as they are of
CDM.

\subsection{Halo Structure}
\label{halos.sec}

\begin{figure*} \begin {center} 
\includegraphics[height=0.45\textheight]{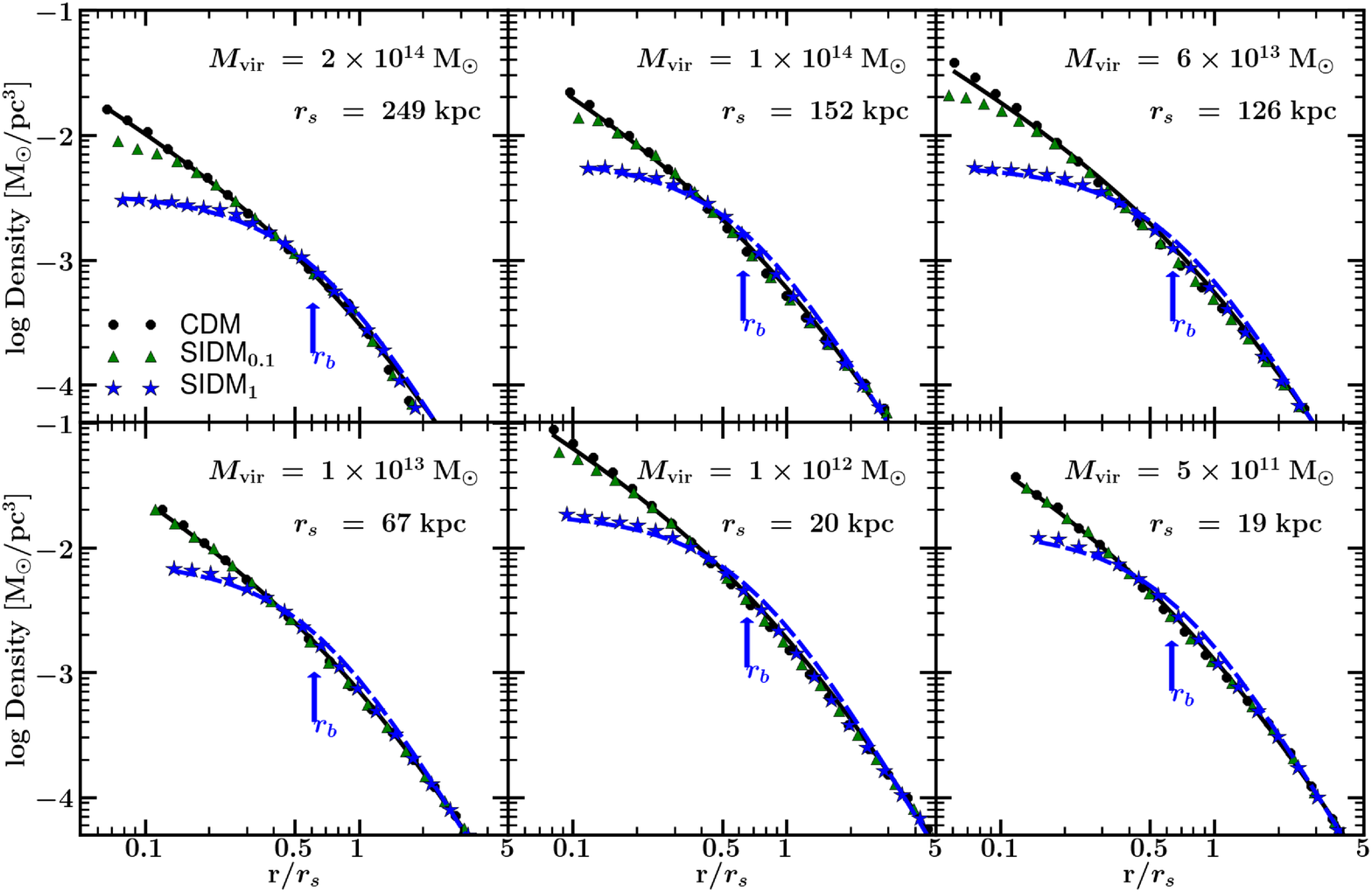}
\end {center} 
\caption{Density profiles for our six example halos from our
SIDM$_1$ (blue stars) and SIDM$_{0.1}$ (green triangles) simulations and their
CDM counterparts. With
self-interactions turned on, halo central densities decrease, forming cored
density profiles. Solid lines are for the best NFW (black) and Burkert (blue)
fits, with the points representing the density at each radial bin found by AHF. 
The arrow indicates the location of the Burkert core radius
$r_\mathrm{b}$. $r_\mathrm{s}$ is the NFW scale radius of the corresponding CDM
halo density profile (black solid line). Burkert profiles provide a reasonable
fit to our SIDM$_1$ halos only because $r_\mathrm{b} \approx r_\mathrm{s}$ for
$\sigmam=1 \cmspg$, so a cored profile with a single scale radius works. As
discussed in \S \ref{analytic.sec} this is not the case for $\sigmam=0.1 \cmspg$
and thus Burkert profiles are not a good fit to our SIDM$_{0.1}$ halos.}
\label{densProfiles.fig} 
\end{figure*}

\begin{figure*} \begin {center} 
\includegraphics[height=0.45\textheight]{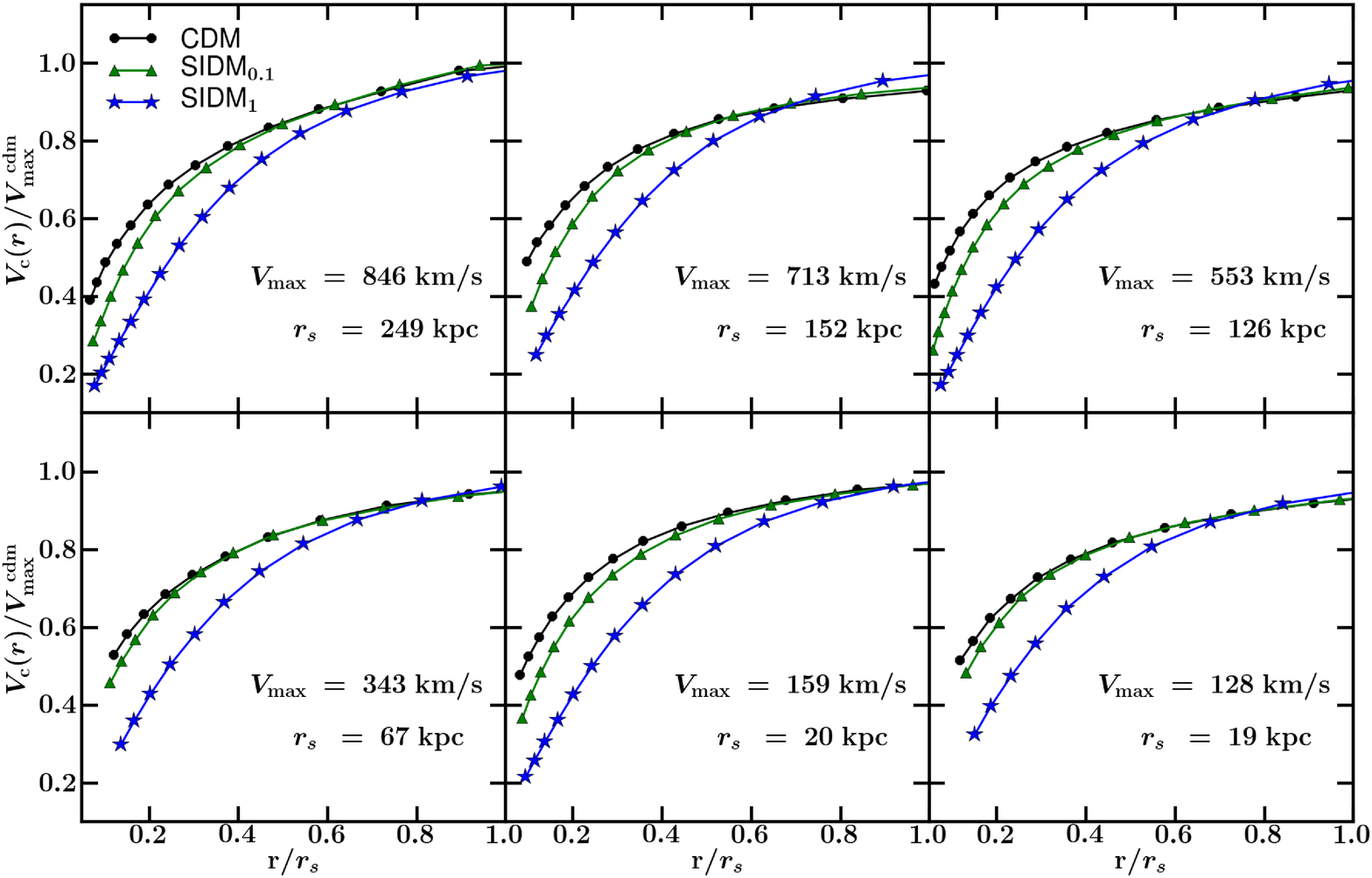}
\end {center} 
\caption{Circular velocity profiles for our example selection of six well
resolved halos from our CDM, SIDM$_1$ and SIDM$_{0.1}$ simulations. The
magnitude of the
circular velocity at small radii $r \lesssim r_\mathrm{s}$ is lowered for all
halos when self-interactions are turned on. $r_\mathrm{s}$ is the NFW scale
radius of the corresponding CDM halo density profile.}
\label{vcircProfiles.fig} 
\end{figure*}

\begin{figure*} \begin {center} 
\includegraphics[height=0.45\textheight]{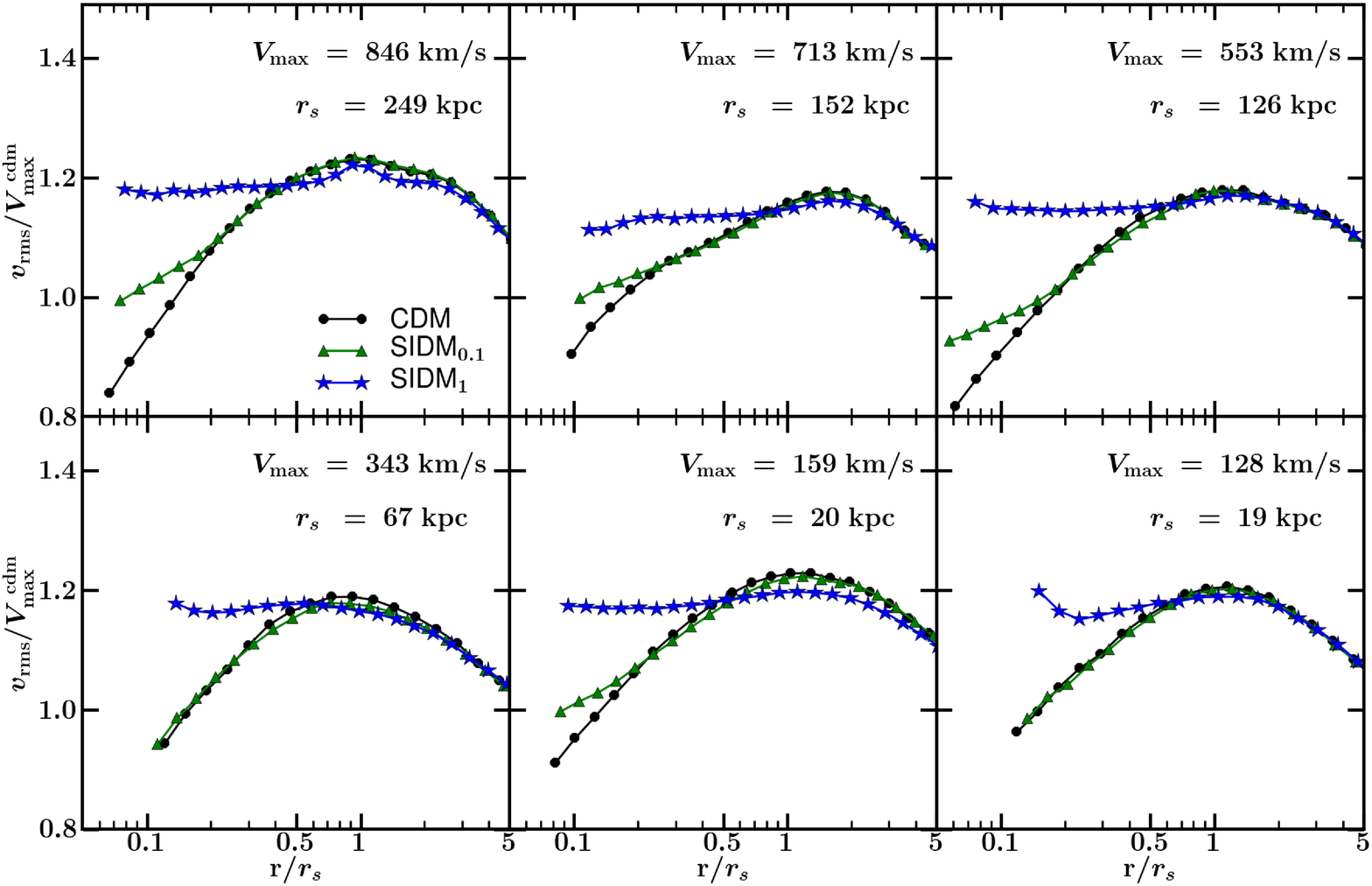}
\end {center} 
\caption{Velocity dispersion profiles for our six example halos from our
SIDM$_1$ and SIDM$_{0.1}$ simulations over-plotted with their CDM counterparts. 
The
velocity dispersion is inflated at small radii and slightly suppressed at large
radii.  The effects
set in at approximately the radius where SIDM particles experience at least one
interaction on 
average over the lifetime of the halo (see Figure \ref{sigvRhoProfiles.fig}).}
\label{vrmsProfiles.fig} 
\end{figure*}

Before presenting statistics on halo structure, we focus on six well resolved
halos that span our full
mass range  $M_{\rm vir} = 5\times 10^{11}-2\times 10^{14} \Msun$, selected from
our full simulation suite, including our two zoom-simulation
halos (Z12 and Z11).    Figures \ref{densProfiles.fig}
through \ref{vrmsProfiles.fig} show radial profiles for the density, circular
velocity and velocity dispersion for all three dark matter cases.
In each figure, black circles correspond to CDM, green triangles to 
 SIDM$_{0.1}$, and blue stars to SIDM$_1$.  All profiles are shown down to the
innermost resolved radius for which the average two-body relaxation time roughly
matches the age of the Universe \citep{poweretal03}.

We begin with the density profiles of halos shown in the six-panel Figure
\ref{densProfiles.fig}.  For each halo in the CDM run we have fit an NFW profile
\citep{nfw97} to its radial density structure:
\begin{equation}
\rho_{\rm NFW}(r) = \frac{\rho_s \, r_s^3}{r(r_s+r)^2},
\end{equation}
 and recorded its corresponding scale radius $r_s$.  The CDM-fit $r_s$ value for
each halo is given in its associated panel along with the halo virial mass.  The
radial profiles for each halo (in both the CDM and SIDM runs) are normalized
with respect to the CDM $r_s$ value in the plot.   This allows our full range of
halo masses to be plotted on identical axes.    
 
 The SIDM versions of each halo show remarkable similarity to their CDM
counterparts at large radii.  However, the SIDM$_1$ cases clearly begin to roll
towards constant-density cores at small radii.  The best resolved halos in the
SIDM$_{0.1}$ runs also demonstrate lower central densities compared to CDM,
though the differences are at the factor of $\sim 2$ level even in our best
resolved systems.  Clearly, higher resolution simulations will be required in
order to fully quantify the expected differences between CDM and SIDM for
$\sigma/m \sim 0.1 \cmspg$.

For the SIDM$_1$ cases we can quantify the halo cores by fitting them to
\citet{burkert1995} profiles
\beq
\label{Burkert.Eq}
\rho_{\rm B}(r) =
\frac{\rho_\mathrm{b}r_\mathrm{b}^3}{(r+r_\mathrm{b})(r^2+r_\mathrm{b}^2)}.
\eeq
These Burkert fits are shown as blue dashed lines.  They are good fits for radii
within $r\sim2-3 \ \rs$, but the quality of the fits gets worse at large radii. 
The blue arrows in each panel show the value of the best-fit Burkert core radius
for the SIDM$_{1}$ halos.  Note that the values are remarkably stable in
proportion to the CDM $r_s$ value at $r_b \simeq 0.7 \, r_s$.

As explained in \S \ref{analytic.sec},  the fact that the SIDM$_1$ profiles are
reasonably well characterized by a single scale-radius Burkert profile may be a
lucky accident, only valid for cross sections near $1 \cmspg$.  It just so
happens that for this cross section the radius where dark matter particles
experience significant scattering sets in at $r \sim \rs$ (see Figure
\ref{sigvRhoProfiles.fig} and related discussion).  For a smaller cross section
(with a correspondingly smaller core) a multiple parameter fit may be necessary.
 Given the beginnings of very small cores we are seeing in the SIDM$_{0.1}$
runs, it would appear that we would need one scale radius to define an $r_s$
bend and a second scale radii to define a distinct core.

Another qualitative fact worth noting is that the density profiles of the
SIDM$_1$ halos overshoot the CDM density profiles near the Burkert core radius
(not as much as the Burkert fits do, but the difference in the data points is
noticeable). This is due to the fact that as particles scatter in the center,
those that gain energy are pushed to larger apocenter orbits. This observation
invites us to consider a toy model for SIDM halos where the effect of SIDM is
confined to a region (smaller than a radius of about $r_b$) wherein particles
redistribute energy and move towards a constant density isothermal core. We will
develop this model further to explain the scaling relations between core size
and halo mass in \S \ref{scaling.sec}.
 
The circular velocity curves for the same set of halos discussed above are shown
in Figure \ref{vcircProfiles.fig}.  The SIDM rotation curves rise more steeply
and have a lower normalization than for CDM within the NFW scale radius $r_s$. 
This brings to mind the rotation curves observed for low surface brightness
galaxies and we will explore this connection later. 
Note though that the peak circular velocity $\Vmax$ actually is slightly higher
for the SIDM$_1$ case because of the mass rearrangement (evident in the density
profiles in Figure \ref{densProfiles.fig}) briefly discussed in the last
paragraph. 
At radii well outside the core radius, the rotation curves of the CDM and
SIDM$_1$ halos converge, though this convergence occurs beyond the plot axes $>
r_s$ for most of the halos shown.

An appreciation of why the density profiles of SIDM halos become cored can be
gained from studying their velocity dispersion profiles compared to their CDM
counterparts, as illustrated in Figure \ref{vrmsProfiles.fig}.  Here $v_{\rm
rms}$ is defined as the root-mean-square speed of all particles within radius
$r$.  While the CDM halos (black) are colder in the center than in their outer
parts (reflecting a cuspy density profile) the SIDM halos have hotter cores,
indicative of heat transport from the outside in.  Moreover, the SIDM halos are
slightly {\em colder} at large radii, again reflecting a redistribution of
energy.  As discussed in the introduction, it is this heat transport that is the
key to understanding why CDM halos differ from SIDM halos in their density
structure \citep{balberg2002,colinetal02,ahn2005,koda2011}. The added thermal
pressure at small radii is what gives rise to the core.   The SIDM$_1$
simulations have sufficient interactions that they have been driven to
isothermal profiles for $r/\rs \lesssim 1$, while for SIDM$_{0.1}$ the $v_{\rm
rms}$ profiles
typically begin to deviate from the CDM lines only at smaller radii, $r/\rs \sim
0.2$, reflecting the relatively lower scattering rate.

The deviations in the SIDM $v_{\rm rms}$ profiles compared to CDM appear to set
in at approximately the radius where we expect every particle to have interacted
once in a Hubble time. This is explored directly in Figure
\ref{sigvRhoProfiles.fig}, where we present a proxy
for the local scattering rate as a function of distance from the halo center:
\begin{equation}
 \rho(r) \, v_{\mathrm{rms}}(r) \propto \Gamma(r) \, (\sigma/m)^{-1}.
 \end{equation}
We have divided out the cross section so it is easier to compare the
SIDM$_{0.1}$ and SIDM$_1$ cases.  Figure \ref{sigvRhoProfiles.fig} presents this
rate proxy in units of $1 \hbox{ Gyr cm}^2/\hbox{g}$: for the SIDM$_1$ case
(with $\sigma/m = 1 \cmspg$) the radius where a typical particle will have
scattered once over a 10 Gyr halo lifetime is  $\rho(r)v_{\mathrm{rms}}(r) =
0.1$.  For the SIDM$_{0.1}$ case (with $\sigma/m = 0.1 \,
\hbox{cm}^2/\hbox{g}$), the ordinate needs to be ten times higher ($\sim 1$) in
order to achieve the same scattering rate.   

By comparing Figure \ref{sigvRhoProfiles.fig} to Figure \ref{vrmsProfiles.fig}
(and to some extent to all Figures 4-6) 
we see that the effects of self-interactions do become evident at radii
corresponding to  $\rho \, v_{\rm rms} \sim 0.1$ for SIDM$_1$ (at $r/\rs \sim
0.8$) and $\rho \, v_{\rm rms} \sim 1$ for SIDM$_{0.1}$ (at $r/\rs \sim 0.2$).  
 Interestingly, for the SIDM$_1$ halos this interaction radius is fairly close
to the Burkert scale radius (shown by the blue arrows).   It should be kept in
mind, however, that the structure of halos can be affected to larger radii
because particles scattering in the inner regions can gain energy and move to
larger orbits. A careful inspection of the density and rotation velocity
profiles shows that this is indeed the case. 

 We will discuss these findings in more detail in Sections \ref{scaling.sec} and
\ref{analytic.sec}.  In particular, in \S \ref{analytic.sec} we present an
analytic model aimed at understanding how the central densities and scale radii
of SIDM halos are set in the context of energetics.  But before moving on to
those issues, we first explore halo substructure in SIDM.

\subsection{Substructure}
\label{subhalos.sec}
The question of halo substructure is an important one for SIDM.  One of the
original motivations for SIDM was to reduce the number of subhalos in the
Milky-Way halo in order to match the relative dearth of observed satellite
galaxies \citep{spergelandsteinhardt00}.  However, the over-reduction of halo
substructure is now recognized as a negative feature of SIDM compared to CDM,
given the clear evidence for galaxy-size subhalos throughout galaxy clusters
\citep{natarajan2009} and the new discoveries of ultra-faint galaxies around the
Milky Way (see \citet{willman2010} and \citet{bullock2010} for reviews). In
fact, one of the most stringent constraints on the self-interaction cross
section comes from analytic subhalo-evaporation arguments
\citep{gnedinandostriker01}. 

Figure \ref{subVmaxFunct.fig} demonstrates that the effects of subhalo
evaporation in SIDM are not as strong as previously suggested on analytic
grounds.  Here we show the cumulative number of subhalos larger than a given 
$\Vmax$ for a sample of well-resolved halos in our CDM (solid), SIDM$_{0.1}$
(dotted), and and SIDM$_1$ (dashed) simulations.  The associated virial masses
for each host halo are shown in the legend.  The left panel presents the $\Vmax$
function for all subhalos within the virial radius of each host and the right
panel restricts the analysis to subhalos within half of the virial radius.   We
see that generally the reduction in substructure counts at a fixed $\Vmax$ is
small but non-zero and that the effects appear to be stronger at small radii
than large.
Similarly, there appears to be slightly more reduction of substructure in the
SIDM cluster halos compared to the galaxy size systems.

\begin{figure*} \begin {center} 
\includegraphics[height=0.45\textheight]
{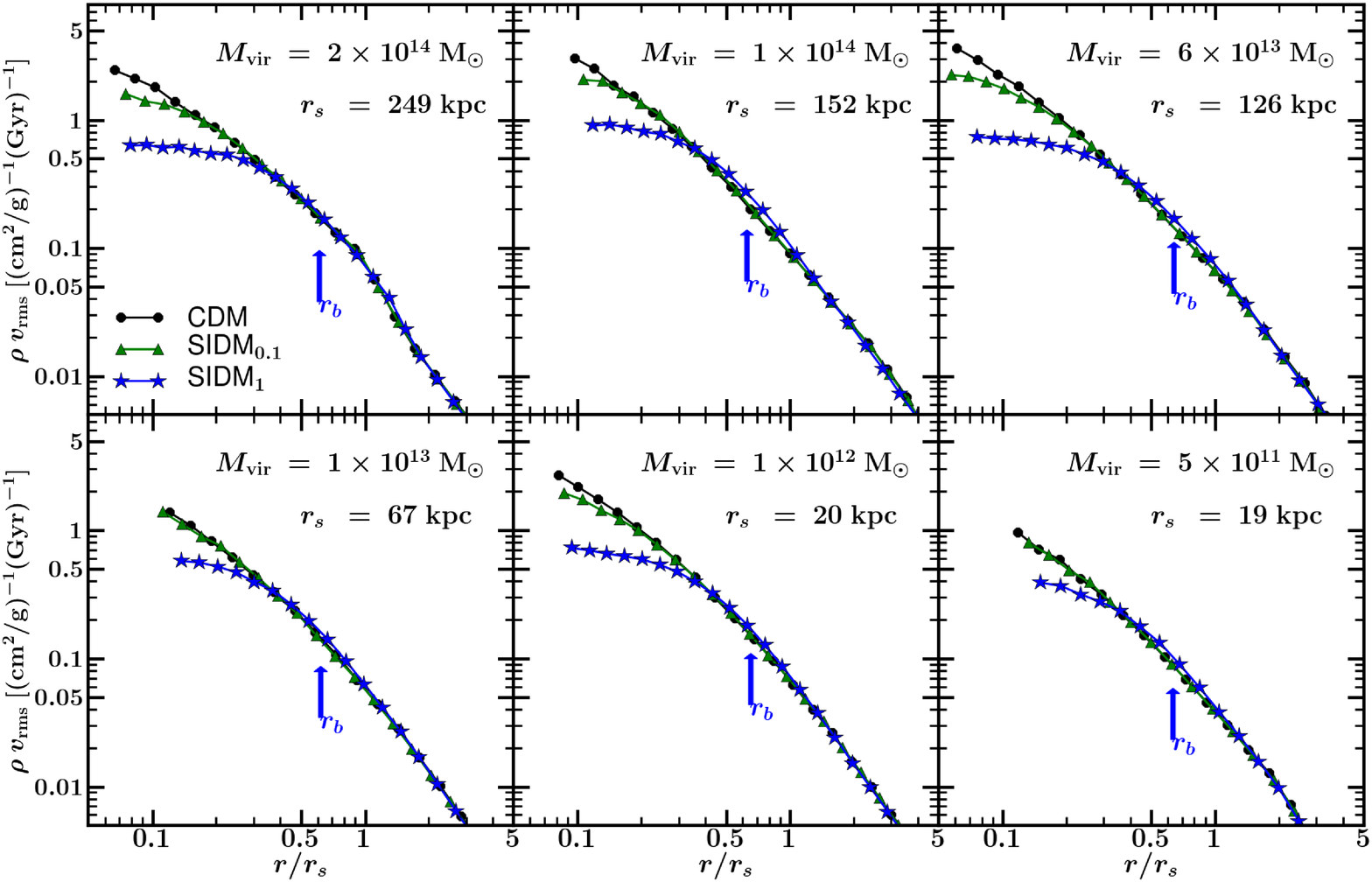}
\end {center} 
\caption{Estimate of the local scattering rate modulo the cross section $\rho
v_{\mathrm{rms}} = \Gamma (\sigma/m)^{-1}$ for six well resolved halos from our
CDM, SIDM$_{0.1}$, and SIDM$_1$ simulations.   The quantity is scaled by $1
\hbox{ Gyr cm}^2/\hbox{g}$, such that $1$ in these units means that each
particle has roughly one interaction per Gyr in SIDM$_{1}$  and $0.1$ per Gyr in
SIDM$_{0.1}$.  
 Based on this argument, the effects of self-interactions in the properties of
halos over $\sim 10$ Gyr should start to become important when the ordinate is
greater than about 0.1 in SIDM$_{1}$ ($r/\rs \sim 0.8$) and greater than about 1
in SIDM$_{0.1}$ ($r/\rs \sim 0.2$).  Comparisons to Figures 4-6 indicate that
this is indeed the case. 
  }
\label{sigvRhoProfiles.fig} 
\end{figure*}

\begin{figure*} \begin {center} 
\includegraphics[width=\textwidth]
{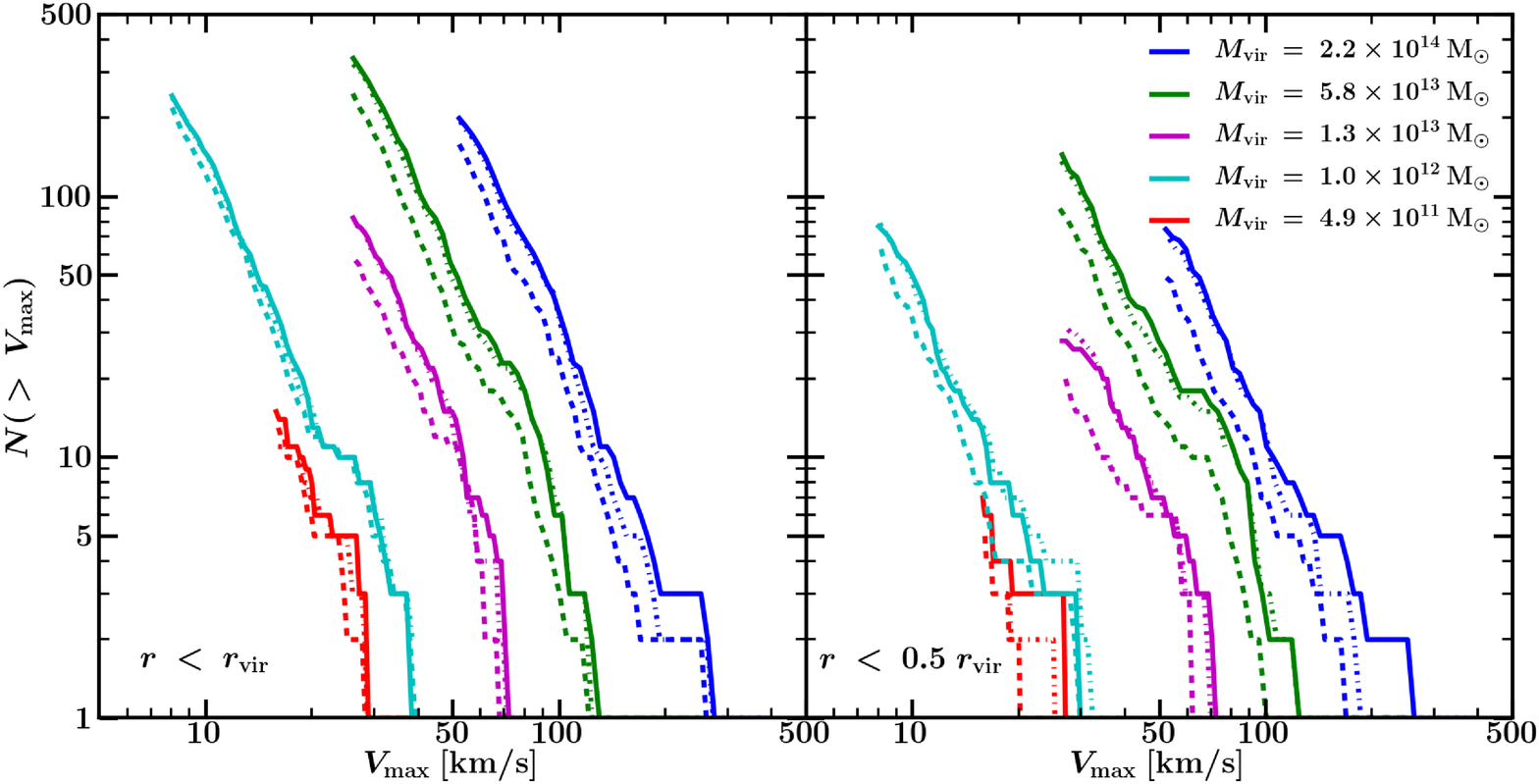}
\end {center} 
\caption{
Subhalo cumulative number as a function of halo peak circular
velocity ($\Vmax$) for several well-resolved halos in
our CDM (solid), SIDM$_{0.1}$ (dotted), and SIDM$_1$ (dashed) simulations.
When looking at all subhalos within $r < \Rvir$ (left), the differences are
small and the slope of the subhalo $\Vmax$ function is the same for the 
CDM and SIDM cases. The offset in the subhalo $\Vmax$ function increases
when we look only at subhalos inside $r < 0.5 \ \Rvir$ (right panel), showing 
that SIDM suppresses the number of subhalos
in the central regions of halos more strongly.}
\label{subVmaxFunct.fig} 
\end{figure*}

We can understand both trends, 1) the increase in the difference between the CDM
and SIDM $\Vmax$ functions as $M_{\mathrm{vir}}$ increases and 2) the increase
in the difference as one looks at the central regions of the halo, using the
results from the previous section as a guide.   The typical probability that
particle in an SIDM subhalo will interact with a particle in the background halo
is 
\begin{equation}
P \approx \langle \rho_{host}(\mathrm{r}) (\sigma/m)
v_{orb}(\mathrm{r})\rangle_{T} \, T, 
\end{equation}
where $v_{orb}(\mathrm{r})$ is the orbital speed of the subhalo at position
$\mathrm{r}$, $\rho_{host}$ is the mass density of the host halo, and $T$ is the
orbital period.  The typical speed of the subhalo is similar to the rms speed of
the smooth component of the halo, and thus $\rho_{host}(\mathrm{r}) (\sigma/m)
v_{orb}(\mathrm{r})$ should be similar to the function we show in Figure
\ref{sigvRhoProfiles.fig}.  At fixed $r/\rs$ we expect $P$ to scale with $\Vmax$
as $\Vmax^3/\Rmax^2$ (given that $\rhos \propto  \Vmax^2/\Rmax^2$), which is a
very mildly increasing function of $\Vmax$ over the range of halo masses we have
simulated.  Note though that we expect scatter at fixed halo mass because of the
scatter in the $\Vmax-\Rmax$ relation \citep{bullock2001}. 

While the increase in destruction of subhalos with host halo mass is not strong,
it is clear from the above arguments that subhalos in the inner parts of the
halo ($r/\rs \ll 1$) should be destroyed but the bulk of the subhalos around
$r/\rs \sim 1$ and beyond should survive for $\sigma/m=1\cmspg$. This effect is
strengthened by the fact that subhalos in the innermost region of the halo were
accreted much longer ago than subhalos in the outskirts, so they have
experienced many more orbits \citep{rochaetal11}.  These arguments explain the
comparisons between the subhalo mass functions plotted in Figure 
\ref{subVmaxFunct.fig}. Our arguments demonstrate that a large fraction of the
subhalos found in CDM halos (most of which are in the outer parts) would still
survive in SIDM halos for $\sigma/m$ values around or below $1\cmspg$.

\vskip 0.2cm  
\noindent Overall in the previous two sections we have seen that the effects of
self-interactions between dark matter particles in cosmological simulations are
primarily in the central regions of dark matter halos, leaving the large scale
structure identical to our non-interacting CDM simulations. Thus we retain the
desirable features of CDM on large scales while revealing different
phenomenology near halo centers.  In the following
section we will move to explore how the properties of SIDM halos presented
here scale with halo mass.

\section{Scaling Relations}
\label{scaling.sec}

In the previous section we saw that while SIDM preserves the
CDM large scale properties of dark matter halos, self-interactions in the
central regions of halos result in a decrease of central densities and the
formation of cores in their density profiles. We found that the
density profiles of halos from our SIDM$_1$ simulations can be relatively well
fit by Burkert density profiles inside $r \sim 2-3 \rs$ (see Figure
\ref{densProfiles.fig}). Here we
define a sample of well resolved halos from all our SIDM$_1$ simulations and
use Burkert fits to their density profiles in order to quantify their central
densities and core sizes. We then provide scaling relations of dark
matter halo properties with maximum circular velocity $\Vmax$. 

The sample of halos used for the rest of this section consists of the two host
halos in our SIDM$_1$-Z11 and SIDM$_1$-Z12 simulations together with the 25
most massive halos from our  SIDM$_1$-50 and the 25 most massive halos from our
SIDM$_1$-25 simulations. That gives us a total of 52 halos spanning a range
$\Vmax = 30-860 \
\kps$  or $\Mvir = 5\times 10^{11} - 2\times 10^{14} \ \Msun$. For this
set of halos the innermost resolved radius, defined by Equation 20 in
\citet{poweretal03}, is always smaller than one third of the
Burkert scale radius from which we define the sizes of cores. It is vital that
we do a conservative comparison to the \citet{poweretal03} radius because both
gravitational scattering and self-interactions lead to the same phenomelogical
result of constant density cores. Most of the halos (other than the 52 we select
here) do not pass this test well enough for the core set by self-interactions to
be resolved with confidence. This desire to be conservative in our presentation
of scaling relations forces us to find these relations from only a small sample
of halos for SIDM$_{1}$ and leaves us with basically no halos to find scalings
for SIDM$_{0.1}$. Also one has to keep in mind that our SIDM$_{1}$ relations
could be biased by selecting only the most massive halos in our full box
simulations. Evidently higher resolution simulations are necessary to find
definitive answers. It is reassuring however that the scaling relations derived
from our analytical arguments in \S \ref{analytic.sec} agree so well with the
ones presented here for $\sigmam = 0.1 \cmspg$.    

We have checked that for all of our halos we resolve the scattering rate out to
at least four times the Burkert scale radius.  Outside of this point 
the scattering rate is underestimated because of our choice of the
self-interaction smoothing length relative to the interparticle spacing (see
\S \ref{test.sec}).  However, the expected scattering rate is negligible with
respect to the Hubble rate outside that radius (Figure
\ref{sigvRhoProfiles.fig}).   Moreover, we have re-run our $50 \hMpc$ boxes for
a range of SIDM smoothing values and found identical results.  Thus we consider
our sample to be well resolved.   

Eight halos in our sample are undergoing significant interactions and have
density profiles that are 
clearly perturbed even in the CDM runs.   We include these eight systems in all
of the following plots
but indicate them with open symbols.  We do not use them in
the best fits for the scaling relations that we provide. 

\begin{figure}
 \begin {center} 
\includegraphics[width=0.5\textwidth]{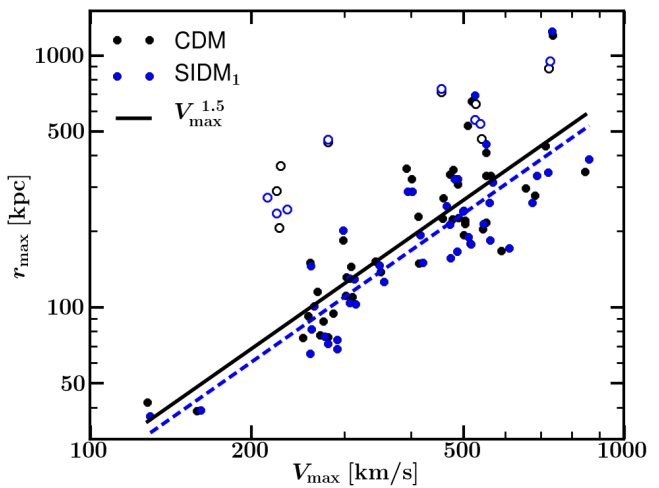}
\end {center} 
\caption{$\Rmax$ vs. $\Vmax$ for our combined sample of well resolved halos
from our SIDM$_1$ and CDM simulations. Open symbols correspond to halos for
which the density profiles showed signs of being perturbed, thus they were not
included in the best fit of the relation. Small differences of about $10\%$
exists in both $\Vmax$ and $\Rmax$, however the slope of $\Vmax$-$\Rmax$
relation is unchanged from CDM to SIDM$_1$.
}
\label{rmaxVmax.fig} 
\end{figure}

We start by examining the global structure of halos as characterized by the
maximum circular velocity $\Vmax$ and the radius where the rotation curve peaks,
$\Rmax$.  The relationship between $\Vmax$ and $\Rmax$ provides a simple,
intermediate-scale measure of halo concentration and we aim to investigate any
differences between SIDM and CDM.
Figure \ref{rmaxVmax.fig} shows the $\Vmax - \Rmax$ relation for CDM (black) and
SIDM$_1$ (blue) halos. We can see that small
differences of about $10\%$ exists in both $\Vmax$ and $\Rmax$, with SIDM$_1$
halos having larger values for $\Vmax$ and smaller for $\Rmax$. This was
already evident in Figure \ref{vcircProfiles.fig}, where the
circular velocity curves of SIDM$_1$ halos seem to peak at slightly smaller
radii and slightly larger velocities than their CDM analogs, even though
SIDM$_1$ curves decrease more steeply at the center. 

The apparent difference is consistent with a picture where energy exchange due
to scattering redistributes the SIDM dark matter particles, with many of the
tightly bound particles scattered onto less bound, high apocenter orbits. Since
the radius at which self-interactions are significant (see Figure
\ref{sigvRhoProfiles.fig}) is smaller than (but close to) $\rs$, it is entirely
reasonable that the scattered particles lead to a new $\Rmax$ for SIDM$_1$ that
is smaller than the CDM $\Rmax$ and a $\Vmax$ that is larger. Notice that the
slope of the $\Vmax-\Rmax$ relation is unchanged from CDM to SIDM$_1$. 
The best-fit relations are: 
\begin{align}\label{RmaxVmax.eq}
\Rmax &= 26.21 \kpc \ \left(\frac{\Vmax}{100 \kps}\right)^{1.45} \ \
(\mathrm{CDM})\,,
\nonumber \\
\Rmax &= 22.46 \kpc \ \left(\frac{\Vmax}{100 \kps}\right)^{1.46} \ \
(\mathrm{SIDM}_1). 
\end{align}

\begin{figure} \begin {center} 
\includegraphics[width=0.5\textwidth]{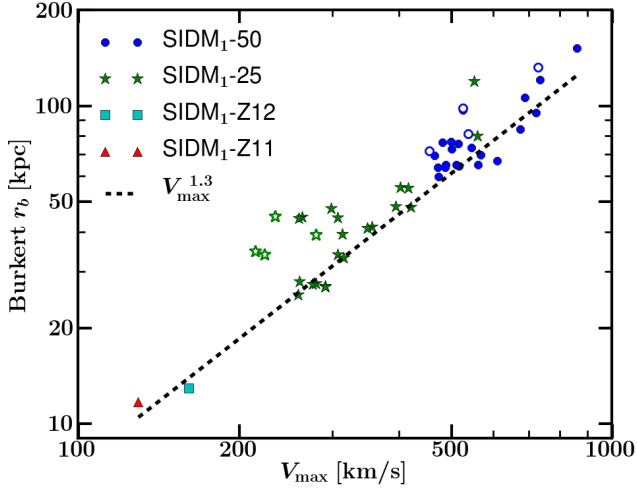}
\end {center} 
\caption{Burkert scale radius vs. $\Vmax$ for our combined sample of well
resolved
halos from our SIDM$_1$-50 (blue circles), SIDM$_1$-25 (green stars),
SIDM$_1$-Z12 (cyan square) and SIDM$_1$-Z11 (red triangle) simulations. Open
symbols correspond to halos that are undergoing mergers.  These perturbed halos
were not included in the fit for the scaling relation. A single
power law holds along the whole range of our sample, suggesting that this
dependence continues towards smaller and larger $\Vmax$ values. }
\label{rbVmax.fig} 
\end{figure}

We continue this discussion by considering the sizes of cores in our SIDM$_1$
simulations as a function of $\Vmax$. The core sizes of halos are quantified by
the scale radius in the Burkert fit to their density profiles, namely
$r_\mathrm{b}$ in Equation \ref{Burkert.Eq}. Figure \ref{rbVmax.fig} shows that
for this relation a single
power law holds along the whole range of our sample. We will come
back to this result in our discussion section (\S \ref{discussion.sec}) on
extrapolating to smaller and larger $\Vmax$ values to make contact with
observations of cores in galaxies and clusters. The power law that best fits our
data is given by
\begin{align}
\label{rb-vmax.eq}
r_\mathrm{b} &= 7.50 \kpc \ \left(\frac{\Vmax}{100 \kps}\right)^{1.31}.  
\end{align}
If we fit to $\Mvir$ instead of $\Vmax$ we get 
\begin{align}
\label{rb-mvir.eq}
r_\mathrm{b} &= 2.21 \kpc \ \left(\frac{\Mvir}{10^{10} \Msun}\right)^{0.43}.  
\end{align}

We note that the scaling with $\Vmax$ is close to that expected for $\Rmax$ or $\rs$. We show this explicitly by fitting for the core size of SIDM$_1$ halos
$r_\mathrm{b}$ as a function of the NFW scale radius $r_\mathrm{s}$ of their CDM
counterparts, as shown in Figure \ref{rbRs.fig}.
We find that the ratio of the core size of a SIDM$_1$ halo
to the scale radius of the corresponding CDM halo varies very mildly with $\Vmax$. In other words, the core sizes are a fixed fraction of the CDM halo scale radius. The relation
that best fits our data is given by
\begin{align}
\frac{r_\mathrm{b}}{\rs} &= 0.71 \ \left(\frac{r_\mathrm{s}}{10
\kpc}\right)^{-0.08}.  
\end{align}\\
This underscores the point that $r_\mathrm{b}$ and $\rs$ are closely tied to
each other and the fact that they are numerically so close to each other is the
reason why a cored profile with a single scale (like a  Burkert profile)
provides a reasonable fit to our SIDM$_1$ halos.
We will explain this striking behavior using an analytic model in the next
section.

The central densities in SIDM$_1$ halos can be defined either as the Burkert
profiles scale density or as the density at the innermost resolved radius. We
have found that both definitions give similar results with no significant
differences. In Figure \ref{rhobVmax.fig}, we show how the Burkert
scale density $\rho_\mathrm{b}$ scales with $\Vmax$. The trend in the
$\rho_\mathrm{b}-\Vmax$ relation is not as strong as for the
$r_\mathrm{b}-\Vmax$ relation, with a scatter as large as about a factor of 3.
We will come back to the implications of this result in our discussion section
(\S \ref{discussion.sec}). The relation that best fits our data is given by
\begin{align}
\label{rhob-vmax.eq}
\rho_\mathrm{b} &= 0.015 \Msun/\pc^3 \ \left(\frac{\Vmax}{100
\kps}\right)^{-0.55}.  
\end{align}
If we fit to $\Mvir$ instead of $\Vmax$ we get 
\begin{align}
\label{rhob-mvir.eq}
\rho_\mathrm{b} &= 0.029 \Msun/\pc^3 \ \left(\frac{\Mvir}{10^{10}
\Msun}\right)^{-0.19}.  
\end{align}
We urge caution when using the above fits to the central densities as it is
likely to be affected by our small sample size given the large scatter. The toy
model discussed in the next section predicts a slightly stronger scaling with
$\Vmax$ . However, the typical densities of order $0.01 \Msun/\pc^3$ for galaxy
halos and $0.001 \Msun/\pc^3$ for cluster halos (see Figure \ref{rhobVmax.fig})
are in line with the predictions of the analytic model.

\bigskip

\noindent In this section we have presented scaling relations for the
properties of halos in our SIDM$_1$ simulations. Our limited resolution
allows us to use only 52 halos spanning a modest mass range,  from which we
throw out eight systems that are undergoing mergers. 
Admittedly, this sample is not large enough to be definitive, especially in
regards to scatter.    However, the strong correlation between the SIDM core
radius $r_\mathrm{b}$ and the counterpart CDM scale radius $\rs$ is clearly
statistically significant and the general trends provide a useful guide for
tentative observational comparisons -- a subject we will return to in the final
section below.

\section{Analytic model to explain the scaling Relations}
\label{analytic.sec}

In this section we develop a simple model to understand the scaling relations
shown in \S \ref{scaling.sec}. This model is based on identifying an appropriate
radius $r_1$ within which self-interactions are effective and demanding that the
mass as well as the average velocity dispersion within this radius is set by the
mass and the average velocity dispersion (within the same radius) of the {\em
same halo in the absence of self-scatterings}. The mass loss due to scatterings 
in the core should be insignificant because particles rarely get enough 
energy to escape and this implies that the mass within
$r_1$ should be close to what it would have been in the absence of
self-interactions. This also implies that the potential outside $r_1$ is
unchanged from its CDM model prediction, but tends to a constant value faster
inside $r_1$. Within this set of approximations, the dominant effect due to scatterings
is to re-distribute kinetic energy in the core, while keeping the total
kinetic energy within $r_1$ the same as it would have had before
self-interactions became important. We have looked at the kinetic energy
profiles in the best-resolved halos in our simulations and have confirmed that
this is indeed a good approximation. Note that in this picture, there is a clear
demarcation of time-scales such that the inner halo structure (say $r \lesssim
\rs$) is set (the same way as in CDM model) well before self-interactions become
important. For cross sections much larger than what we are interested in here,
this need not hold.  

To set up the model, we start by recalling that self-interactions work to create
an isothermal core (see Figure \ref{vrmsProfiles.fig}) that is isotropic (both
spatially and in velocity space). Using the spherical Jeans equation, one can
then see that for a system with these properties
\begin{equation}
v_{\rm rms,0}^2 = 3\sigma_{r}(0)^2=2\pi \xi^{-1} G \rho(0) r_0^2\,,
\end{equation}
where we have defined $r_0$ to be the expansion parameter such that
$\rho(r)\sigma_r(r)^2 = \rho(0)\sigma_r(0)^2(1-\xi(r/r_0)^2)$ when $r\ll r_0$,
and $\sigma_r$ is the radial velocity dispersion. The form of the Taylor
expansion for $\rho(r)\sigma_r(r)^2$ is dictated by the Jeans equation for
density profiles that tend to a constant value, as may be readily ascertained by
taking the derivative of $\rho(r)\sigma_r(r)^2$. To fix $r_0$, we will choose it
to be equivalent to the Burkert scale radius where the density is one-fourth of
the central density. The parameter $\xi$ encapsulates uncertainties from the
profile and velocity dispersion anisotropy in the outer parts of the halo. We
test various models and find that a range of 2-3 for $\xi$ is largely consistent
with most parameterizations and hence we fix $\xi=2.5$. If we specify the
central velocity dispersion, then with an additional constraint on the core
region ({\em i.e.}, $r_1$), we would be able to back out both the core radius
and the core density. 

We then set $v_{\rm rms,0}^2$ equal to the average velocity dispersion squared
({\em i.e.}, two times kinetic energy divided by mass) within the region $r_1$
in the absence of self-interactions. This basically demands that the kinetic energy within
$r_1$ is unchanged from the value it would have had in the absence
of self-interactions. Note, however, that we are setting the average velocity
dispersion squared equal to $v_{\rm rms,0}^2$ and not the corresponding average 
in the SIDM halo. This is an approximation, but one that is degenerate with choosing the
$\xi$ parameter.

To finish specifying this model, we need a density profile for the region inside
$r_1$. A Burkert profile has a velocity dispersion profile (assuming isotropy)
that asymptotes very slowly to the central dispersion. For small radii, the
radial dispersion profile is slowly increasing (with radius) because of the
$r/r_\mathrm{b}$ term in the Taylor expansion for the density profile. If we
want a flatter central dispersion profile (as is observed for the SIDM$_1$
halos), we can fix this by either assuming an isothermal profile or something
like $1/(1+1.52(r/r_0)^2)^{3/2}$. The final results turn out to be qualitatively
similar for these profiles. Hence we adopt a Burkert profile for ease of
comparison to the fits presented here and then check the results with more
appropriate profiles later. Our two constraints (on the radial velocity
dispersion and mass) fully specify the density and radial scales of the Burkert
profile. 

In order to obtain scaling relations we need to estimate $r_1$, which demarcates
the inner region where self-interactions are effective from the outer region
that is mostly undisturbed by the self-interactions. In reality, this divide
will not be sharp but we will see that the main features of the scaling
relations are well-captured by this simple model. We define $r_1$ to be the
region where each particle on average suffers one interaction.  Since the region
outside is assumed to be unperturbed by interactions, we may estimate $r_1$ as:
\begin{equation}
\Gamma(r_1) t_{\rm age}=1.3 \rho_{\rm CDM}(r_1) v_{\rm rms,CDM}(r_1)
\frac{\sigma}{m} t_{\rm age}=1\,,
\label{gamma1.eq}
\end{equation}
where we set age ($t_{\rm age}$) to be 10 Gyr for now, keeping in mind that
larger halos have a shorter age and that major mergers can reset the timer. 
We will consider what happens when $t_{\rm age}$ is a function of halo mass
shortly. The factor 1.3 is 
$\langle \left | \vec{v}-\vec{u}\right | \rangle/\sqrt{\langle v^2\rangle}$ 
for a Maxwellian distribution where $\vec{u}$ and $\vec{v}$ are the velocities of the
two interacting dark matter particles. We have not attempted to use a more realistic 
velocity distribution since the dependence of this factor on a possible high-velocity
cut-off to the distribution function was found to be fairly mild. 

For the density profile in the absence of self-interactions, we assume
a NFW profile and to fix the velocity dispersion we use the observed fact that
the phase space density is a power-law in radius \citep{taylor2001}. By noting
that $v_{\rm rms,CDM}(r) = (\rho_{\rm CDM}(r)/Q(r))^{1/3}$ and using a
phase-space density profile $Q(r) = Q(r_\mathrm{s}) (r/r_\mathrm{s})^{-\eta}$
\citep{taylor2001,rasia2004,ascasibar2004,dehnen2005,ascasibar2008}, we may
fully specify the dependence of $r_1$ on the cross-section and halo parameters
(say $\Vmax$ and $\Rmax$). For the phase-space  density profile we use a
power-law index $\eta=2$ and $Q(r_\mathrm{s})=0.3/(G \Vmax \Rmax^2)$ derived
from jointly fitting our relaxed CDM halos; these parameters are very similar
to the fits provided in \citet{ascasibar2008}. 

Let us first look at how $r_1$ scales with $\rs$ in the NFW density profile. One
notes that $\rhos =1.72  \Vmax^2/(G \Rmax^2)$ and hence $\rhos \Vmax \propto
\Vmax^3/\Rmax^2$ which is a very mildly increasing function of $\Vmax$ as our Equation \ref{RmaxVmax.eq} shows. Thus Equation \ref{gamma1.eq} implies that $r_1/\rs$ should be roughly a constant. Numerically,
we find that $r_1/\rs \simeq 0.7-0.8$ over the range of $\Vmax$ of interest for $\sigmam =1 \cmspg$. 

Having now specified $r_1$, we are ready to look at the scalings of
$r_\mathrm{b}$ and $\rho_\mathrm{b}$. For our assumed value of $\xi$, $v_{\rm
rms,0}^2\simeq 2.5 G \rho_\mathrm{b} r_\mathrm{b}^2$. Thus we are looking for
the value of $r_\mathrm{b}/r_\mathrm{s}$ that solves,
\begin{equation}
\left\langle \frac{\rho_\mathrm{s}}{(r/r_s)(1+r/r_s)^2}
\frac{(r/r_s)^{\eta}}{Q(r_\mathrm{s})} \right \rangle(r_1)= (v_{\rm rms,0})^3\,,
\label{constraint.eq}
\end{equation}
with the constraint that $M_b(r_1)=M_\mathrm{NFW}(r_1)$ where
$M_\mathrm{NFW}(r)$ and $M_\mathrm{b}(r)$ are the masses enclosed within radius
$r$ for NFW and Burkert profiles, respectively. We note that if
$r_\mathrm{b}/\rs$ is not a strong function of $\Vmax$ and since we know
$r_1/\rs$ is a mild function of $\Vmax$, then the mass constraint essentially
sets $\rho_\mathrm{b}r_\mathrm{b}^3/(\rhos\rs^3)$ to be  a constant or
$\rho_\mathrm{b}r_\mathrm{b}^3 \propto (\rhos\rs^3)$. This implies $v_{\rm
rms,0}^2 r_\mathrm{b} \propto \Vmax^2 \Rmax$. Now Equation \ref{constraint.eq} sets
$v_{\rm rms,0} \simeq \Vmax$ because $r_1/\rs$ is a mild function of $\Vmax$ and
it therefore follows that $r_\mathrm{b} \propto \rs$ is a consistent solution to
the above equations. As a check we note that assuming $r_1/\rs=0.7-0.8$ gives
$v_{\rm rms,0} \simeq 1.1 \Vmax$, in reasonable agreement with our SIDM$_1$ simulation
results (see Figure \ref{vrmsProfiles.fig}). This simple model thus  predicts
that $r_\mathrm{b}/r_\mathrm{s}$ should not vary much with $\Vmax$ in agreement
with the observed scaling relations from the SIDM$_1$ simulation. 

\begin{figure} \begin {center} 
\includegraphics[width=0.5\textwidth]{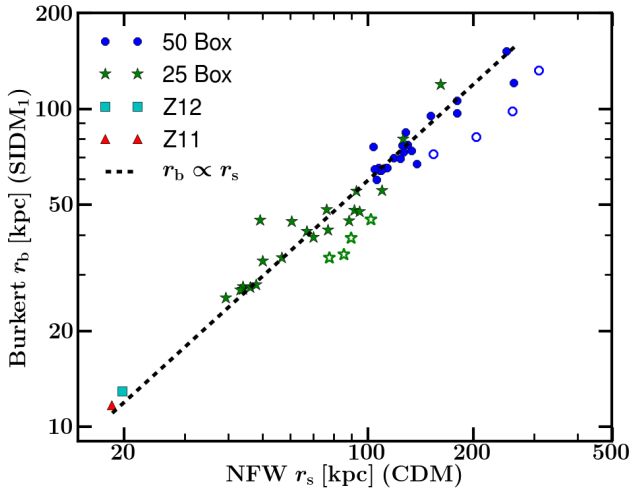}
\end {center} 
\caption{Burkert scale radius in SIDM$_1$ halos vs. the NFW scale radius in
their CDM counterparts. Points and labels are the same as in Figure
\ref{rbVmax.fig}. There is a one-to-one correlation indicating that the core
size of SIDM$_1$ halos scales the same as the scale radius of CDM halos with
$\Vmax$ }
\label{rbRs.fig} 
\end{figure}

In detail, the model predicts that $r_\mathrm{b}/\rs=0.5-0.6$ for dwarf to
cluster halos in good agreement with the fits to our SIDM$_1$ halos, but 
about 25\% smaller for $\Vmax \sim 100\, \rm km/s$. It departs
from the results of the simulation in predicting that $r_\mathrm{b}/\rs$
increases gently with $\Vmax$, whereas Figure \ref{rbRs.fig} predicts that this
ratio should decrease gently with $\Vmax$. We find that this departure from
simulations is likely related to the assumption of a constant age for all halos.
To generalize our model, we use the results of \citet{wechsler2002} who show
that the virial concentrations of halos are correlated with their formation
times, and in particular $c_{\rm vir}=4.1 (1+z_{\rm form})$ for a particular
definition of formation time. We invert this equation to derive an estimate of
the halo age using $z_{\rm form}$. With the age thus specified in
Equation~\ref{gamma1.eq}, we find that now $r_\mathrm{b}/\rs$ decreases gently
with $\Vmax$ in substantial agreement with the fit to our simulations. Thus the
reason that larger halos have a smaller $r_\mathrm{b}/\rs$ is because
self-interactions have had less time to operate. We note that the values for the
core radius in the analytic model with halo mass dependent $t_{\rm age}$ are 
uniformly about 25\% smaller, but this should not be a cause for concern given 
the approximation in demanding a sharp transition at $r_1$. 

\begin{figure} \begin {center} 
\includegraphics[width=0.5\textwidth]{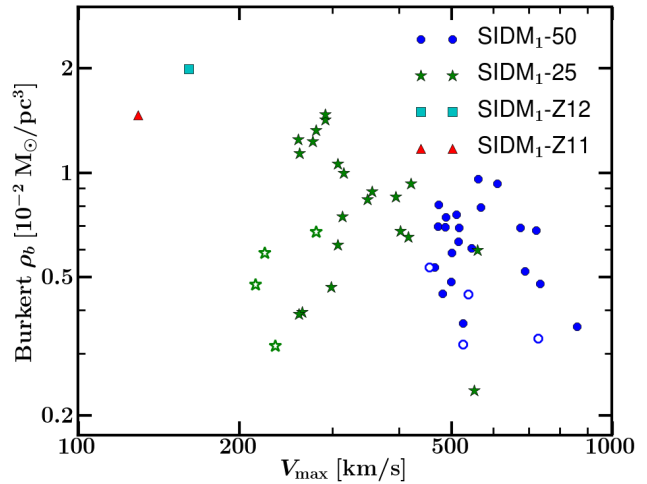}
\end {center} 
\caption{Burkert scale density vs. Vmax. Points and labels are the same as in
Figure\ref{rbVmax.fig}. The trend in the $\rho_\mathrm{b}-\Vmax$ relation is
not as clear as for the $r_\mathrm{b}-\Vmax$ relation, with a scatter of
up to a factor of 3.}
\label{rhobVmax.fig} 
\end{figure}

Given the Burkert core radius $r_\mathrm{b}$ and the central velocity 
dispersion $v_{\rm rms,0}$, one can easily check that the central density
$\rho_\mathrm{b}$ is about $0.01 \Msun/\pc^3$ for $\Vmax=300 \kps$ halos and
$0.005 \Msun/\pc^3$ for  $\Vmax=1000\kps$ in this analytic model. These 
numbers and the scaling with $\Vmax$ for $\rho_\mathrm{b}$ (when including 
the halo mass dependent $t_\mathrm{age}$) are in good agreement with the 
densities in Figure \ref{rhobVmax.fig} and the fit in Equation \ref{rhob-vmax.eq}. 
As we have indicated before, the scaling relation 
for the central density should be interpreted with care given the large scatter.
Given the tight correlation between core radius and $\rs$, it is possible that
the substantial scatter in the central density arises in large part due to the
scatter introduced by the assembly history in the concentration-mass relation.
This has important implications for fitting to the rotation velocity profiles of
low-surface brightness spirals \citep{kuzioetal10} and deserves more work.

The simple model constructed above also provides insight into the core collapse
time scales. In particular, as long as the outer part (region outside $r_1$) dominates the
potential well and sets the average central temperature (or the total kinetic energy in
the core), we do not expect core collapse. This is simply because core collapse requires 
uncontrolled decrease in temperature, which is prohibited here. Once $r_1$ moves out 
well beyond $\Rmax$ or to the virial radius, there is significant loss of particles and core 
collapse may occur if there are no further major mergers. The time scale for this process 
is much longer than the age of the universe for $\sigma/m=1\cmspg$ because the inner 
core is at $r_1 < \rs$ after 10 Gyr for this self-interaction strength and we see no evidence 
for significant mass loss.

\section{Observational Comparisons}
\label{discussion.sec}

The goal of this section is to discuss our results in comparison to
observationally inferred properties of dark-matter density profiles.  In
particular, we will focus on the core densities and core sizes.  \S
\ref{discussion_predict.sec} presents our expectations for SIDM$_1$ and
SIDM$_{0.1}$.  Our predictions for $\sigma/m = 1\hbox{ cm}^2/\hbox{g}$ are
anchored robustly to our simulations, though they do require some extrapolation
beyond the mass range directly probed by our simulations  ($\Vmax = 130 - 860 \
\kps$).   For $\sigma/m = 0.1 \hbox{ cm}^2/\hbox{g}$ the predictions are much
less secure because the associated core sizes are of order our resolution limit,
thus we rely on our our analytic model more directly here.
In \S \ref{discussion_obs.sec}, we discuss our predictions in light of
observations of dark-matter halos for a wide range of halo masses.  In \S
\ref{discussion_subhalo.sec}, we discuss our results on subhalos in the context
of past work and constraints on SIDM based on subhalo properties.

Before proceeding with this discussion we would like to clarify how we quantify
core sizes.  In this work, we have fit the $\sigma/m=1 \cmspg$ halos with
Burkert density profiles.  However, many observational constraints on cores on
galaxy scales come from fitting pseudo-isothermal density profiles with core
size $r_{\mathrm{pi}}$ to data \citep[e.g.,][]{simonetal05,kuzioetal08},
although some constraints do come from Burkert modeling \citep{saluccietal12}. 
We found that pseudo-isothermal density profiles
also give good fits to the inner regions of the SIDM$_1$ halos, but
Burkert fits are better because of that profile's $\rho \propto r^{-3}$
dependence at large radii.  
For a pseudo-isothermal density
profile ($\propto 1/(r_c^2+r^2)$), the density decreases to 
one-fourth the central density at 1.73 times its core radius $r_c$. 
Thus, as a crude approximation, one may convert the Burkert radius 
to the equivalent pseudo-isothermal core radius by multiplying by
a factor of 0.58 ($r_c \simeq r_b/1.73$).

\subsection{Predicted Core Sizes and Central Densities in
SIDM}\label{discussion_predict.sec}

\subsubsection{SIDM with $\sigmam = 1 \ \cmspg$.}

The central properties of dark-matter halos have been inferred from observations
from tiny Milky Way dwarf spheroidal (dSph) galaxies ($\Vmax \lesssim 50\hbox{
km}/\hbox{s}$) to galaxy clusters ($\Vmax \gtrsim 1000\hbox{ km}/\hbox{s}$).  If
we extrapolate the results from our set of SIDM$_1$ simulations using
Eqs. (\ref{rb-vmax.eq})-(\ref{rhob-mvir.eq}) we predict that SIDM halos with
$\sigmam = 1 \cmspg$ would have the
following (Burkert) core sizes and central densities:

\bigskip

\noindent For galaxy clusters $(\Vmax \simeq 700-1000\hbox{ km}/\hbox{s})$:
\begin{equation}
r_\mathrm{b} \simeq (95-155) \kpc  \: ; \:
\rho_\mathrm{b} \simeq  (0.005-0.004)  \Msun \pc^{-3} \nonumber 
\end{equation}

\noindent For low-mass spirals $(\Vmax \simeq 50-130
\kps)$:
\begin{equation}
r_\mathrm{b} \simeq (3-10) \kpc \, ;  \, \rho_\mathrm{b} \simeq  (0.02-0.01)  
\Msun \pc^{-3} \nonumber 
\end{equation}

\noindent For dwarf spheroidals galaxies $(\Vmax \simeq 20-50 \kps)$:
\begin{equation}
r_\mathrm{b} \simeq (0.9-3) \kpc  \,   ;  \, \rho_\mathrm{b} \simeq \:
(0.04-0.02)  \Msun  \pc^{-3} \nonumber 
\end{equation}

\bigskip 

\noindent Although we can not completely determine the scatter in our scaling
relations
due to low number statistics, it is important to note from Figs.
\ref{rbVmax.fig} and \ref{rhobVmax.fig} that a scatter of at
least a factor of 2 in core sizes, and at least a factor of 3 in central
densities, is expected for a given $\Vmax$.   We suspect that these differences
are in large part a result of the diversity of merger histories of dark-matter
halos.  Note that the $\Vmax$-$\Rmax$ and $Q(\rs)$ scalings assumed in the analytic model are the median values. The strong dependence of the SIDM halo profiles on these quantities makes it clear that the scatter in these relations will introduce significant scatter in the halo core sizes and core densities. Thus the analytic model should also provide a simple way to understand (some of the) scatter seen for SIDM$_1$ halo properties. In future work we will characterize the relation between the core
properties and merger history in the context of a detailed discussion of the scatter in the
scaling relations, especially on scales that we do not resolve with our current
simulations. 

\subsubsection{SIDM with $\sigmam = 0.1 \ \cmspg$}

As discussed in \S \ref{halos.sec} our SIDM$_{0.1}$ simulations are not well
enough resolved to definitively measure a
core radius for any of our halos, much less define scatter in that quantity. 
Nevertheless, our best resolved systems do demonstrate
some clear deviations from CDM and allow us to cautiously estimate individual
core densities.  Referring back to Figure 4, we see that in our two best
resolved cluster halos (at $\Mvir \simeq 10^{14} \Msun$)  the SIDM$_{0.1}$ core
densities approach $\sim 0.01 \, \Msun/\pc^3$ -- each at least a factor of $\sim
3$ denser than their SIDM$_1$ counterparts.    Similarly, in our Z12 Milky-Way
case, the SIDM$_{0.1}$ core density appears to be approaching $\sim 0.1 \,
\Msun/\pc^3$ compared to $\sim 0.02 \, \Msun/\pc^3$ in the SIDM$_1$ case.

Given the lack of well-resolved halo profiles, it is worth appealing to the
analytic model presented in \S \ref{analytic.sec} to estimate core radii for
SIDM$_{0.1}$.
Using exactly the same arguments (including the halo mass dependent age), we find
that $r_1/\rs \simeq 0.05-0.12$ in the 100-1000 km/s $\Vmax$ range and 
a corresponding Burkert core radius $r_b/\rs \simeq 0.09-0.17$.  We note that the Burkert 
radius is close to but slightly larger than $r_1$. It is important to keep in mind that 
in this analytic model we are only explicitly fitting the inner "self-interaction zone" of $r<r_1$. 
This does not imply that the entire halo has to be well-fit by the Burkert profile. 
Recall that a single-scale Burkert profile only works as well as it does for
$\sigma/m = 1\hbox{ cm}^2/\hbox{g}$ because $r_b \approx r_s$, such that to a
good approximation there is only one relevant length scale. For the smaller
cross section that we are now considering we expect the core and NFW scale radii
to be widely separated, suggesting that a generic functional form for SIDM halos
should have two scale radii.  A wide separation between the SIDM$_{0.1}$ core
and $r_s$ does appear to be consistent with the highest resolved halos presented
in Figure 4. However, we note that given the strong correlation
between $r_\mathrm{b}/\rs$, we still expect a one-parameter family of models for
a given $\sigmam$. 

To see how dependent our results are on the shape of the {\em inner} halo profile, we modify
the analytic model to include a density profile that decreases with radius as
$1/(1+(r/r_c)^2)^{\alpha/2}$. For this density profile, the velocity dispersion
profile has the right form to match our simulation results. The price we pay is
the introduction of a new parameter $\alpha$. We set this parameter alpha by
additionally demanding that the slope of the mass profile (i.e., density) is
continuous at $r_1$, so that the mass profile joins smoothly with the NFW mass
profile. This picks out a narrow range $\alpha=5.5-7.0$ as the solution over most
of the $\Vmax$ range of interest (with smaller values corresponding to lower $\Vmax$). 
Interestingly, this implies that at $r_1$, the
slope of the density profile is very close to $-2$ for the entire range of
$\Vmax$ values of interest.  Note that while the mass profile is continuous, the
slope of the density profile is not matched smoothly at $r_1$ (since the slope
of the NFW profile would be closer to $-1$ at $r_1 \ll \rs$) . This probably
signals that if the matching were not done sharply (at $r_1$), the density
profile of SIDM would overshoot that of CDM and catch up at some radius beyond
$r_1$ (as is seen in the comparison of SIDM$_1$ and CDM density profiles). 

As a check we apply this $\alpha$-model to $\sigmam =1 \ \cmspg$ case and find
that the results are qualitatively the same as the model with the Burkert profile. The quantitative 
differences are at 20\% level with the densities being smaller and inferred Burkert core radii (where 
density is 1/4 of the central density) larger compared to the Burkert profile model. The
predicted slope of the density profile at $r_1$ is close to $-2.5$ implying
a smoother transition to the NFW profile (since $r_1 \sim \rs$ for $\sigmam = 1\
\cmspg$), as is seen Figure \ref{densProfiles.fig}. 

For the $\sigmam =0.1 \ \cmspg$ case, we obtain $r_c/\rs=0.08-0.17$ and an
equivalent Burkert core radius (where the density is one-fourth of the central
density) $r_\mathrm{b}/r_s=0.06-0.14$, in substantial agreement with the results 
we obtained using the Burkert profile.   Thus our analysis would suggest that core sizes 
 $\sim 0.1\rs$ for $\sigmam=0.1\ \cmspg$.   The results
from the analytic model for $\sigmam=0.1\ \cmspg$ also seem consistent with our
simulation results; see Figure 6 where the $v_{\rm rms}$ profiles for
SIDM$_{0.1}$ start to deviate from CDM at $\sim 0.2 r_s$. 

Based on the discussion above we conclude that for $\sigmam = 0.1 \ \cmspg$ we
expect: 
\bigskip

\noindent For galaxy clusters $(\Vmax \simeq 700-1000 \hbox{ km}/\hbox{s})$:
\begin{equation}
r_\mathrm{b} \sim (16-20)  \kpc \: ; \:
\rho_\mathrm{b} \sim 0.04 \Msun \pc^{-3} \nonumber 
\end{equation}

\noindent For low-mass spirals $(\Vmax \simeq 50-130
\kps)$:
\begin{equation}
r_\mathrm{b} \sim (0.6 - 2.5) \kpc \,  ;  \, \rho_\mathrm{b} \sim  0.2-0.1  \Msun
\pc^{-3} \nonumber 
\end{equation}

\noindent For dwarf spheroidals galaxies $(\Vmax \simeq 20-50 \kps)$:
\begin{equation}
r_\mathrm{b} \sim (0.2-0.6) \kpc   \,  ;  \, \rho_\mathrm{b} \sim  0.5-0.2  
\Msun  \pc^{-3} \nonumber 
\end{equation}

\bigskip 

\noindent These values do not include the scatter from mass assembly history. It
is probably reasonable to assume a factor of 2 scatter for both core radii and
core densities based on what we see in SIDM$_1$. It is also possible that the
core densities are $\sim 50\%$ smaller than what we would see in simulations, given
that the SIDM$_1$ simulations have core densities that are somewhat larger than the
predictions from the analytic model. For the dwarf spheroidal galaxies, the
values should be interpreted with caution as it is the prediction for field
halos with $\Vmax$ range $20-50 \kps$. 

While these values are somewhat tentative compared to those presented above for
SIDM$_{1}$ (given our lack of direct simulation fits), two factors are
reassuring. First, the analytic model is based on the simple assumption that
scattering redistributes kinetic energy within the inner halo and the
non-trivial aspect of the model is defining this "inner halo" region. There is
no reason to suspect that this assumption or the prescription breaks down for
SIDM$_{0.1}$ halos when it works so well in describing the SIDM$_1$ halos. The
predicted densities are in line with those inferred for the best resolved halos
in our SIDM$_{0.1}$ simulations
(shown in Figure 4 and discussed above).  For the core radii, we reiterate that
the label ``$r_{\rm b}$" should be interpreted (according to its definition in
the analytic model)  as the radius where the density reaches one-fourth the
asymptotic core density. The overall profile of a halo with such a small core
compared to $\rs$ will not be fit by the Burkert form.  Note that the
strong correlations we predict between the core radius and the NFW scale radius
raise the intriguing possibility that the SIDM halos may be also well fit (modulo scatter) by a
single parameter profile as is the case for CDM. 

Next, we compare our predictions for SIDM core properties against data and show 
that the core radii and densities appear to be consistent
with that seen in real data, motivating future
simulations with high enough resolution to resolve cores in SIDM$_{0.1}$ halos.

\subsection{Observed Core Sizes and Central Densities vs.
SIDM}\label{discussion_obs.sec}
In this section, we explore the predictions for the properties of density
profiles with SIDM in the context of observational constraints on density
profiles.  We also revisit previous constraints on SIDM from observations in
light of our simulation suite.

\subsubsection{Clusters}

One of the tightest SIDM constraints from the first generation of SIDM studies
emerged from one cluster simulation and one observed galaxy cluster. 
Specifically, 
\citet{yoshida00} simulated an individual galaxy cluster with different SIDM
cross sections.  When comparing the core size of this simulated cluster to the
core size estimated by \citet{tysonetal98} for CL 0024+1654, they found that the
observed core in CL 0024+1654 would be consistent with SIDM only if $\sigma/m
\lesssim 0.1\hbox{ cm}^2/\hbox{g}$.  Since that time, evidence has emerged that
this particular cluster is undergoing a merger along the line of sight
\citep{czoske2001,czoske2002,zhang2005,jee2007,jee2010,umetsu2010}.  Thus, this
cluster is not the ideal candidate for SIDM constraints based on the properties
of relaxed halos, and the \citet{yoshida00} constraint is not valid in this
context.
 
 Using X-ray emission, weak lensing, strong lensing, stellar kinematics of the
brightest cluster galaxy (BCG) or some combination thereof, the mass
distributions within a number of galaxy clusters have been mapped in the past
decade.
  \citet{arabadjisetal02} placed a conservative upper
limit of $75 \kpc$ on the size of any constant-density core, and an average
density within the inner $50 \kpc$ of $\sim 0.025 \Msun/\pc^3$ for an halo with
an estimated mass $M\sim 4 \times 10^{14} \Msun$.  

\citet{sandetal04}, \citet{sand2008} , 
\citet{newmanetal09},  and \citet{newmanetal11} all find central density
profiles in clusters shallower than the NFW CDM prediction.  
The difference in the work between these authors and others is that they use
stellar kinematics of the BCG to constrain the density profile of the cluster
dark-matter halo on small scales.  While this probe of the density profile is
more sensitive on small scales than strong lensing is, proper inference of the
dark halo properties depends on accurate modeling of the BGC density profile and
equilibrium structure.   They have typically assumed a  ``gNFW'' profile in
order to constrain the central densities: $\rho(r) \propto 1/(x^g (1+x)^{3-g})$
with $r=x\rs$ and the NFW form obtained when $g=1$.
 The \citet{newmanetal09} and \citet{newmanetal11} mass
models of $M\sim10^{15} \Msun$ clusters show average dark matter central
densities within $10$ kpc of $\sim 0.03-0.06 \Msun/\pc^3$ and $r_s$ of order 100
kpc.  Note that 10 kpc is typically the smallest radii our simulations can
resolve.

\citet{saha2006} and \citet{saha2009} studied the mass structure of 3
cluster halos from gravitational lensing and obtained density profiles
that are consistent with $\rho \propto r^{-1}$ outside the inner $10-20 \kpc$
regions. Similarly \citet{morandi2010} and \citet{morandi2012} find that the
radial mass distribution of cluster dark-matter halos are consistent with NFW
predictions outside 30 kpc in projection.  The CLASH multi-cycle treasury
program on the \emph{Hubble Space Telecope} is finding many new strongly lensed
galaxies in about a set of 25 massive clusters \citep{postman2012}.  Initial
results from this program show that the total density profile of these clusters
(or total density minus the brightest cluster galaxy), if modeled as spherically
symmetric, are consistent with NFW predictions for the halo alone if the gNFW
functional form is used in the fit \citep{zitrin2011,coe2012,umetsu2012}. 
However, \citet{morandi2010} and \citet{morandi2011a} find that spherical mass
modeling of galaxy clusters typically results in an overestimate of the the
cuspiness of the density profile, although axially-symmetric modeling is found
to lead to underestimates \citep{meneghetti2007}.  Thus, the present status of
the density profiles of the CLASH clusters is unclear and clearly an interesting
data set to look forward to.

We note here a complexity involved in using the lensing results to constrain
SIDM models. Lensing provides mass in cylinders along the line of sight and this
2D mass profile is sensitive to mass from a large range of radii. As an example,
lets consider mass within 30 kpc in projection. If we were to do something
extreme and create a zero density core inside $30 \kpc$ sphere, the differences
in the 2D mass profile would be less than a factor of 2 for clusters in the
$10^{14-15} \Msun$ mass range. For SIDM$_{0.1}$, the differences are
comparatively benign. Our analytic model predicts that differences relative to
CDM at about $0.1 \rs$ (which is $10-40 \kpc$ for $10^{14-15} \Msun$ virial mass
range) are 20-30\%, which implies SIDM$_{0.1}$ surface mass density profiles are
very similar to CDM on these scales. But for SIDM$_1$ the expected differences
would be measurably large. 

On a related technical note, we discourage the use of the gNFW functional form
when thinking about models that deviate from the CDM paradigm.  In the SIDM
case, for $\sigma/m < 1\ \cmspg$, there will generically be two scale radii: one
is the NFW-like scale radius which is the result of hierarchical structure
formation \citep{lithwick2011}, and the second is the core radius from
dark-matter self-scattering.  For $\sigma/m = 1 \ \cmspg$, as we explained in
detail in \S \ref{analytic.sec}, the two scales are about the same.  If most
of the cluster data constrain the density profile beyond a SIDM core, as they
may for weak lensing and X-ray studies, the gNFW or NFW fit is dominated by
those data, and a core will not be ``detected'' in the fit.  
In future work, we will simulate halos with a broader range of $\sigmam$ and
provide SIDM-inspired density profiles to the community.

\begin{figure} \begin {center} 
\includegraphics[width=0.5\textwidth]{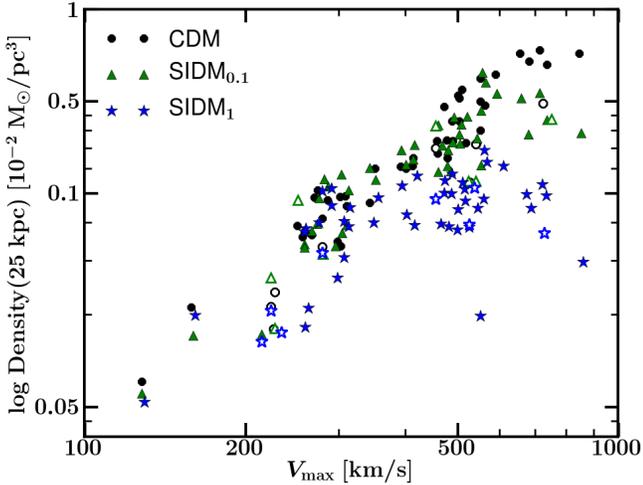}
\end {center} 
\caption{ Dark matter average density within 25 kpc  vs. $\Vmax$  for resolved
halos in our  CDM, SIDM$_{0.1}$, and SIDM$_1$ simulations. It is clear that
SIDM$_1$ has significantly lower densities than CDM halos at group and cluster
scales. For the SIDM$_{0.1}$ model, the differences are muted and only appear on
cluster scales. Thus observations of central densities in clusters likely
provide the most promising avenue to look for signatures of SIDM with cross
sections in the vicinity of $0.1 \ \cmspg$.}
\label{rho25Vmax.fig} 
\end{figure}

The results discussed above seem to suggest that the density profile beyond
about 25 kpc should be close to the predictions from the NFW profile. To test
this we plot the average physical density within 25 kpc for well-resolved halos
in our CDM (black), SIDM$_{0.1}$ (green), and SIDM$_1$ (blue) simulations in
Figure \ref{rho25Vmax.fig}. We see that for the most massive halos, the $\sigmam
= 1 \ \cmspg$ run produces densities at $25$ kpc that are $\sim 2-3$ times lower
than their CDM counterparts. Thus it seems like the measured densities in
clusters rule out $\sigmam=1\ \cmspg$ SIDM model.  At the same time the $\sigmam
= 0.1 \ \cmspg$ simulations are quite similar to CDM at these these radii,
though beginning to show some differences as we discussed earlier in this
section.  Analyses that combine information from X-rays, lensing and BCG stellar
kinematics seem to suggest lowered densities (e.g., \citet{newmanetal11}) that
would be compatible with SIDM$_{0.1}$. Given this outlook, it is reasonable to
conclude that estimates of the central dark matter density in clusters will
provide essential tests of interesting SIDM models. 

\subsubsection{Low-Mass Spirals}

For low-mass spirals with maximum circular velocities
in the range $50-130 \kps$, constant-density cores with sizes of
$\sim 0.5-8 \kpc$ and central densities of approximately $\sim 0.01-0.5
\Msun/\pc^3$ have been observed \citep{debloketal01, simonetal05,
sanchezsalcedo05, kuzioetal08, kuzioetal10, ohetal11a, saluccietal12}. Similar
to what we found for clusters scales, SIDM with $\sigmam = 1 \ \cmspg$ would
be able to reproduce the {\em largest} core sizes observed in low-mass galaxies
but it predicts central densities that are too low.  
SIDM with $\sigmam = 0.1 \ \cmspg$ would be much more consistent.  Moreover, the
predicted log-slope of the density profile at 500 parsecs for $\sigmam = 0.1 \
\cmspg$ halos in the 50-130 $\kps$ range is $-0.5$ to 0, both facts consistent
with results from THINGS \citep{ohetal11a}. Note that the slope at 500 pc for
the $\sigmam =1 \ \cmspg$  model is 0 in the same $\Vmax$ range, which is not
consistent with the scatter seen in the data.

We conclude, as before, that the observed densities and core radii are not
consistent with SIDM$_{1}$ but are fairly well reproduced in SIDM models with
$\sigmam \simeq 0.1 \ \cmspg$.

\subsubsection{Dwarf Spheroidals in the Milky Way Halo}

The least massive and most dark-matter-dominated galaxies provide an excellent
setting to confront the predictions of different dark matter models with
observations. Recent work by \citet{mbketal11a, mbketal11b} has found that the
estimated central densities of the bright Milky Way dwarf spheroidal satellites
are lower than the densities of the massive subhalos in dark-matter-only
simulations.  
SIDM offers a way to solve this problem because it reduces the central density
of halos. Thus in SIDM, the massive subhalos \emph{do} host the luminous dSph
but have shallower density profiles than predicted in CDM simulations.  This has
recently been demonstrated by \citet{vogelsberger12}. We do not directly compare 
to \citet{vogelsberger12} because their work is focused on the subhalos of the
Milky Way and the velocity-independent cross section that they simulate 
($\sigmam = 10 \cmspg$) is larger than the cross sections considered in 
our work. 

Regardless of whether Milky Way dSphs have cuspy or cored dark-matter halos, we
may estimate the enclosed mass, and hence average density, around the half-light
radius of the stellar distribution.  Mass estimates within $300 \pc$ and mass
profile modelings using stellar kinematics together with chemo-dynamically
distinct stellar subcomponets of Milky Way dwarf spheroidal galaxies suggest
central densities of approximately  $\sim 0.1 \Msun/\pc^3$ 
 \citep{strigarietal08, wolfetal10,
walkerandpenarrubia11, amoriscoandevans12, wolfandbullock12}.  For the faintest
dSph Segue 1, the density within the half-light radius (about 40 pc) is measured
to be about $2.5^{+4.1}_{-1.9}\Msun/\pc^3$ \cite{Simon2011,Martinez2011}. The
errors on Segue 1 density are large but it is clear that if SIDM is to
accommodate this result, it must allow for large scatter in the core sizes and
densities for small $\Vmax$ halos. With a factor of 2-3 scatter in the densities
quoted earlier for SIDM$_{0.1}$ halos, Segue 1 would appear to be compatible
with SIDM$_{0.1}$ if its $\Vmax$ value is towards the lower end of the $20-50
\kps$ range in $\Vmax$.

For the two dSph galaxies that appear to have cored density profiles (Fornax and
Sculptor), the cores sizes must be of order$\sim 0.2-1 \kpc$
\citep{walkerandpenarrubia11}. For small halos
with circular velocities in the $20-50 \kps$ range, which is close to the
expected peak circular velocities of dwarf spheroidal halos before infall into
the Milky Way host halo, an SIDM with $\sigmam = 1 \ \cmspg$ predicts core sizes
in the order of $\sim0.8-3.0 \kpc$, with central densities of about
$\sim0.02-0.04 \Msun/\pc^3$. Therefore, we find again that $\sigmam = 1 \
\cmspg$ cannot reproduce the observed high central densities.  On the other
hand, our estimates suggest that an SIDM model with $\sigma/m = 0.1\ \cmspg$
would produce central densities and core sizes consistent with the Milky Way
dSph.\\

\noindent In this last section we have used the analytic results that explain
the scaling relations for the
core sizes and central densities of halos in our SIDM$_1$ and SIDM$_{0.1}$
simulations, to
extrapolate our results to scales ranging from galaxy clusters to dwarf
spheroidal galaxies and to lower cross sections. We have found that $\sigmam = 1
\ \cmspg$ would be unable to reproduce the observed high central densities. 
Remarkably, we find that the observations should be consistent with the
predictions of a self-interacting dark matter with cross section in the ballpark
of $\sigmam = 0.1 \ \cmspg$. These expectations are based on the scaling
relations seen in SIDM$_1$ simulations and our analytic model, which is
consistent with the results from our direct $\sigmam = 0.1 \ \cmspg$ simulations
at the radii where we can trust our simulations. This deserves further study
both in terms of simulations with SIDM cross section values smaller than $1 \
\cmspg$ and more detailed comparisons to observations. Our current look at the
global data does not suggest a need for a velocity dependent cross section as
has been previously suggested. In the companion paper (Peter, Rocha, Bullock and
Kaplinghat, 2012) we show that these SIDM models are also consistent with
observations of halo shapes. 

\subsection{Observed Substructure vs. SIDM}\label{discussion_subhalo.sec}
In Figure \ref{subVmaxFunct.fig} we showed that the number of subhalos for
$\sigma/m = 1 \ \cmspg$ is not significantly different from CDM predictions,
especially in galaxy-scale halos.  This is interesting because it means that
SIDM fails to deliver on one of the original motivations for considering this
model of dark matter.  Recall that \citet{spergelandsteinhardt00} originally
promoted SIDM as a solution to the missing satellites problem
\citep{klypin1999,moore1999}, stating that many subhalos would be evaporated by
interactions with the background halo.  Given the new discoveries of ultra-faint
galaxies around the Milky Way and the high likelihood of many more discoveries
from surveys like LSST \citep{willman2010,bullocketal2010}, a significant reduction
in substructure counts may very well be a negative characteristic of any non-CDM
model \citep{tollerud2008}.

However, in Milky Way mass halos, SIDM with  $\sigma/m = 1\ \cmspg$ will yield a
significant probability for subhalo particle scattering only for systems that
pass within  $\sim 10\hbox{ kpc}$ of the host halo center. Thus, for this cross
section, we can form interesting-sized cores but largely leave the subhalo mass
function unaffected in Milky Way-mass halos.  For smaller cross sections, the
differences between SIDM and CDM subhalo mass functions will be even smaller. 
We note that we are not the first to find that SIDM can form cores but not solve
the missing satellites problem; it was first discussed in \citet{donghia2003}.

This finding is also interesting in the context of other alternatives to CDM. 
Warm dark matter (WDM) models, for which the outstanding difference from CDM is
that dark-matter particles have high speeds at matter-radiation equality and a
related free-streaming cutoff in the matter power spectrum, predict a
suppression in the halo (and subhalo) mass function at small scales.  Otherwise,
the abundance and structure of halos and subhalos is nearly indistinguishable
from CDM \citep{villaescusa2011,maccio2012}.  WDM halos may be less concentrated
than CDM halos on scales not much larger than the free-streaming scale, but are
still \emph{cusped}.  They are only significantly cored right at the
free-streaming scale, at which the halo and subhalo abundance is highly
suppressed.  Thus, each of the two leading modifications to CDM can solve only
one of the two historical motivations for looking beyond the CDM paradigm.

The lack of subhalo suppression for $\sigma/m \lesssim 1 \ \cmspg$ has
implications for another of the SIDM halo constraints from a decade ago. 
\citet{gnedinandostriker01} set a constraint excluding the range of $0.3 <
\sigma/m < 10^4 \ \cmspg$ based on the fundamental plane of elliptical galaxies.
 The argument rests on the observation that there are not significant
differences in the fundamental plane of field ellipticals and cluster
ellipticals \citep[e.g.,][]{kochanek2000,bernardi2003,labarbera2010}. Elliptical
galaxies have a significant amount of dark matter within their half-light radii,
with more massive ellipticals having larger mass-to-light ratios, either caused
by varying stellar mass-to-light ratios or varying dark matter content
\citep{padmanabhan2004,Tollerud2011,Conroy2012}. \citet{gnedinandostriker01}
argue that elliptical galaxies falling into cluster-mass halos should have dark
matter evaporated from their centers if $\sigma/m \neq 0$, which would cause the
stars in the elliptical galaxy to adiabatically expand and hence move the galaxy
off the fundamental plane.   

However, in our simulations, we find that few subhalos are fully evaporated, and
that the subhalo $\Vmax$ function is not greatly different for
$\sigma/m = 1 \ \cmspg$ from CDM.  In addition, our analytic arguments show that
the trend with (host) halo mass for the evaporation of subhalos at fixed $r/\rs$
is mild. This suggests that the \citet{gnedinandostriker01} constraints are
overly conservative even at the $\sigma/m \simeq 1 \ \cmspg$ level.  The main
caveats are that the suppression of the subhalo $\Vmax$ function is higher in
more massive clusters and that the suppression is highest at the center of the
cluster halo.    It would also be interesting to see if there are any
differences in the fundamental plane as a function of projected distance in the
cluster, both observationally and in simulations.  For all of these reasons, it
would be worthwhile to perform simulations of elliptical galaxies in clusters
with SIDM and explore the fundamental-plane constraints in more depth.

\bigskip 

To summarize, although we have not fully resolved the cores of $\sigmam=0.1
\cmspg$ SIDM halos, the intuition gleaned from our analytic model (tested agains
the SIDM$_{1}$ results) and our moderately-resolved simulation results suggest
that $\sigmam = 0.1 \cmspg$ is an excellent fit to the data across the range of
halo masses from dwarf satellites of the Milky Way to clusters of galaxies.
Values of cross section over dark matter particle mass in this range are fully consistent with the
{\em published} Bullet cluster constraints (cf. \S\ref{intro.sec}), 
measurements of dark matter density on small-scales
and subhalo survival requirements. In a companion paper (Peter, Rocha, Bullock
and Kaplinghat 2012), we show that this model is also consistent with halo shape
estimates. It is therefore important to simulate galaxy and cluster halos with
cross sections in the $0.1 \cmspg$ range. 

\section{Summary and Conclusions}
\label{sumandconc.sec}

We have presented a new algorithm to include elastic self-scattering of dark
matter particles in N-body codes and used it to study the structure of
self-interacting dark matter (SIDM) halos simulated in a full cosmological
context.  Our suite of simulations (summarized in Table 1) rely on identical
initial conditions to explore SIDM models with velocity-independent cross
sections $\sigma/m = 1 \ \cmspg$ and $\sigma/m = 0.1 \  \cmspg$  as well as a
comparison set of standard CDM simulations (with $\sigma/m = 0$).  

Our primary conclusion is that while SIDM looks identical to CDM on large
scales, SIDM halos have constant density cores, with core radii that scale in
proportion to the standard CDM scale radius ($r_{\rm core} \simeq \epsilon \,
\rs$).    The relative size of the core increases with
increasing cross section ($\epsilon \simeq 0.7$ for $\sigma/m = 1$ and $\epsilon
\sim 0.2$ for $\sigma/m = 0.1 \ \cmspg$).
Correspondingly, at fixed halo mass, core densities decrease with increasing
SIDM cross section.  For both core radii and core densities, there is significant 
scatter about the scaling with $\Vmax$ of the halo. The scaling relationship is strong 
enough that measurements of dark 
matter densities in the cores of dark matter dominated galaxies and large galaxy
clusters likely provide the most robust constraints on the dark matter cross
section at this time.  In a companion paper (Peter, Rocha, Bullock and
Kaplinghat, 2012) we demonstrate, contrary to previous claims, that SIDM
constraints from halo shape measurements may be less restrictive than (or at
least similar to those from) measurements of absolute core densities alone.  

Based on our simulation results we conclude that the dark matter self-scattering
cross section must be smaller than $1 \ \cmspg$ in order to avoid
under-predicting the observed core densities in galaxy clusters, low surface
brightness spirals (LSBs), and dwarf spheroidal galaxies.  However, an SIDM
model with a {\em velocity-independent} cross section of about $\sigma/m = 0.1 \
 \cmspg$ appears capable of reproducing reported core sizes and central
densities of dwarfs, LSBs, and galaxy clusters.   Higher resolution simulations
with better statistics will be needed to confirm this expectation.   

\bigskip

\noindent An accounting of our results are as follows:

\begin{itemize}
\item
Outside of the central regions of dark matter halos ($r \gtrsim 0.5 R_{\rm
vir}$)  the large scale properties of SIDM cosmological simulations are 
effectively identical to CDM simulations.  This implies that all of the
large-scale confirmations of the CDM theory apply to SIDM as well. \\

\item
The subhalo $\Vmax$ function in SIDM with $\sigma/m = 1 \ \cmspg$ differs by
less than $\sim 30\%$ compared to CDM across the mass range $5\times
10^{11}M_\odot - 2\times 10^{14}M_\odot$ studied directly with our simulations .
Differences in the $\Vmax$ function with respect to CDM are only apparent deep
within the centers of large dark-matter halos.  Thus, although is possible, it
will be difficult to constrain SIDM models based on the effects
subhalo evaporation.  \\

\item
SIDM produces halos with constant density cores, with correspondingly lower
central densities than CDM halos of the same mass.
For $\sigmam = 1 \ \cmspg$, our simulated halo density structure is reasonably
well characterized by a Burkert (1995) profile fit with a core size $r_{\rm b}
\simeq 0.7 \rs$, where
$\rs$ is the NFW scale radius of the same halo in the absence of
self-interactions.  Core densities tend to increase with decreasing halo mass
($\rho_b \propto \Mvir^{-0.2}$) but demonstrate about a factor of $\sim 3$
scatter at fixed mass (likely owing to the intrinsic scatter in dark matter halo
concentrations). \\

\item SIDM halo core sizes, central densities, and associated scaling relations
can be understood in the context of a simple analytic model. The model treats
the SIDM halo as consisting of a core region, where self-interactions have
redistributed kinetic energy to create an approximately isothermal cored
density profile; and an outer region, where self-interactions are not effective. The
transition between these regions is set by the strength of the self-interactions
and this model allows us to make quantitative predictions for smaller cross
sections where the cores are not resolved by our simulations.  Based on this
model and a few of our best resolved simulated halos we find 
core sizes $\sim 0.1 \rs$ for $\sigmam = 0.1 \ \cmspg$. \\

\item
Halo core densities over the mass range from $10^{15} - 10^{10} \Msun$ in SIDM
with $\sigmam = 1 \ \cmspg$  are too low  ($ \sim 0.005 - 0.04 \, \Msun/\pc^3$)
to match observed central densities in galaxy clusters ($\sim 0.03 \Msun/\pc$)
and dwarf spheroidals ($\sim 0.1 \Msun/\pc^3$).  \\

\item Halo core central densities in SIDM with $\sigmam = 0.1 \ \cmspg$ are in
line with those observed from galaxy clusters to tiny dwarfs  ($0.02 - 0.5
\Msun/\pc^3$) without the need for any velocity dependence.  The densities are
more consistent with observations than those predicted in dissipationless CDM
simulations, which are generically too high. SIDM models with this cross section
over dark matter particle mass value are consistent with 
Bullet cluster observations, subhalo survival requirements and, as we show in a 
companion paper (Peter, Rocha, Bullock and Kaplinghat, 2012), measurements of 
dark matter halo shapes.

\end{itemize}

Future work is necessary to expand both the dynamic range of our simulations in
halo mass and resolution as well as the dynamic range in cross sections.  These
simulations are necessary in order to make detailed comparisons with
observations given the exciting possibility that dark matter self-interaction
with $\sigmam$ in the ballpark of $0.1 \ \cmspg$ could be an excellent fit to
the central densities of halos over 4-5 orders of magnitude in mass.

\section*{Acknowledgments}
MR was supported by a CONACYT doctoral Fellowship and NASA grant NNX09AG01G.
AHGP is supported by a Gary McCue Fellowship through the Center for Cosmology at
UC Irvine, NASA Grant No. NNX09AD09G at UCI, National Science Foundation (NSF) grant 0855462 at UC Irvine, and the NSF under Grant No. NSF PHY11-25915 while visiting the Kavli Institute
for Theoretical Physics. JSB was partially supported by the Miller Institute for
Basic Research in Science during a Visiting Miller Professorship in the
Department of Astronomy at the University of California Berkeley. JO was supported by a
Fullbright-MICINN Postdoctoral Fellowship.  MK is supported by NASA grant NNX09AD09G and NSF grant 0855462. This research was supported in part by the Perimeter Institute of Theoretical Physics during a visit by MK. Research at Perimeter Institute is supported by the Government of Canada through Industry Canada and by the Province of Ontario through the Ministry of Economic Development and Innovation. 
The work of LAM was carried out at Jet Propulsion Laboratory, California Institute of Technology, under a contract with NASA. LAM acknowledges NASA ATFP support.
Simulations were performed in the Pleiades supercomputer of the NASA Advanced Supercomputing (NAS) Division, and the Kraken supercomputer of the National Institute for Computational Sciences
(NICS) through an XSEDE allocation.

\bibliography{mybibs}
\appendix
\onecolumn

\section{Derivation of The Hard-Sphere Interaction Rate in N-body simulations}
\label{appendixA}
The challenge is to represent a microphysical scattering process in a
macroscopic context in which neither a fluid nor collisionless treatment is
appropriate.  In order to develop a Lagrangian technique in which to represent
the scattering process, we start with the Boltzmann equation.  Particles with
mass $m$, a hard-sphere scattering cross section $d\sigma/d\Omega$ (as a
function of center-of-mass scattering angle), and a distribution function
$f(\mathbf{x},\mathbf{v},t)$ evolve as
\begin{align}\label{eq:boltzmann}
\frac{Df(\mathbf{x},\mathbf{v},t)}{Dt} &= \Gamma[ f,\sigma ] \\
  &= \int d^3\mathbf{v}_1 \int d\Omega \frac{d\sigma}{d\Omega} \left|
\mathbf{v} - \mathbf{v}_1 \right| \left[ f(\mathbf{x},\mathbf{v}^\prime,t)
f(\mathbf{x},\mathbf{v}_1^\prime,t) - f(\mathbf{x},\mathbf{v},t)
f(\mathbf{x},\mathbf{v}_1,t)\right]. \label{eq:Boltz}
\end{align} 
Here, $D/Dt$ is a Lagrangian time derivative and $\Gamma[f,\sigma]$ is the
collision operator.  If the particles were collisionless, the Lagrangian
time-derivative of the distribution function would be zero; the phase-space
density of particles would be conserved.  The left-hand expression in the
brackets in Equation (\ref{eq:Boltz}) represents scattering of particles into a
small
patch of phase space centered on $(\mathbf{x},\mathbf{v})$, and the right-hand
expression (after the $-$ sign) represents scattering out of that patch of phase
space.  If $\mathbf{v}$ and $\mathbf{v}_1$ represent the initial velocities of
the primary and target particles, then $\mathbf{v}^\prime$ and
$\mathbf{v}_1^\prime$ are their post-scatter velocities, which are related to
the initial velocities by the center-of-mass scattering angle $\Omega$.

The key step in being able to represent the scattering process in a simulation
is the ansatz that the evolution of the coarse-grained distribution function
$\hat{f}$ (the distribution function averaged over several times the
interparticle spacing) is a good representation of the evolution of the
fine-grained distribution function $f$.  In other words, the ansatz is that the
solution to 
\begin{align}
  \frac{D\hat{f}}{Dt} = \int d^3\mathbf{v}_1 \int d\Omega
\frac{d\sigma}{d\Omega} \left| \mathbf{v} - \mathbf{v}_1 \right| \left[
\hat{f}(\mathbf{x},\mathbf{v}^\prime,t)
\hat{f}(\mathbf{x},\mathbf{v}_1^\prime,t) - \hat{f}(\mathbf{x},\mathbf{v},t)
\hat{f}(\mathbf{x},\mathbf{v}_1,t)\right]\label{eq:coarseBoltz}
\end{align}
is the same as the solution for $f$ in Equation (\ref{eq:Boltz}) averaged over a
patch of phase space.  If this is the case, our next step is to discretize
Equation
(\ref{eq:coarseBoltz}) such that we can solve the Boltzmann equation by Monte
Carlo N-body methods.

To discretize Equation (\ref{eq:coarseBoltz}), we consider a particle-based
Lagrangian method in which each particle in the N-body simulation represents a
patch of phase space.  In the absence of collisions, the simulation particles trace out
geodesics in the gravitational field of the particles.  When we discretize the
phase space, we do it as follows:
\begin{align}
  \hat{f}(\mathbf{x},\mathbf{v},t) = \sum_i (M_i/m) W(|\mathbf{x} -
\mathbf{x}_i|; h_i) \delta^3(\mathbf{v} - \mathbf{v}_i).\label{eq:fdiscrete}
\end{align}
Here, $i$ labels a discrete macro particle representing a patch of phase space
that has mass $M_i$ ; 
thus each macro particle represents a patch of phase space
inhabited by $M_i/m$ of the true particles.  We assume a delta-function form for
the velocity distribution because each macro particle travels at only one speed.
 We treat each macro particle as being smoothed out in configuration space with
a smoothing kernel $W$ with smoothing length $h_i$.  The reason for treating
each macro particle as inhabiting a finite region of configuration space is that
we want the local estimate of the density
\begin{align}
  n(x) = \int d^3 \mathbf{v} \hat{f}
\end{align}
to be smooth.  Preliminary tests show that smoothness is necessary to properly
estimate the collision term of the Boltzmann equation.  Note that in the main
text we
 use $M_i = m_p$ and $h_i = \hsi$ for all $i$.  This is because all of the
N-body particles have the same mass in our simulations and
 we have fixed $\hsi$ to be constant for all particles in the simulations we
present.

In the particle-based discretization of the Boltzmann equation, the fact that
each particle represents a patch of phase space means that we must discretize
the collision operator; we must integrate the collision operator over the patch
of phase space inhabited by a single particle.  Thus, if a specific particle
represents a patch of phase space of size $\delta \mathbf{x}_p \delta
\mathbf{v}_p$, we must calculate 
\begin{align}
\int_{\delta \mathbf{x}_p} d^3 \mathbf{x} \int_{\delta \mathbf{v}_p} d^3
\mathbf{v} \frac{D\hat{f}}{Dt} = \int_{\delta \mathbf{x}_p} d^3 \mathbf{x}
\int_{\delta \mathbf{v}_p} d^3 \mathbf{v} \int d^3\mathbf{v}_1 \int d\Omega
\frac{d\sigma}{d\Omega} \left| \mathbf{v} - \mathbf{v}_1 \right| \left[
\hat{f}(\mathbf{x},\mathbf{v}^\prime,t)
\hat{f}(\mathbf{x},\mathbf{v}_1^\prime,t) - \hat{f}(\mathbf{x},\mathbf{v},t)
\hat{f}(\mathbf{x},\mathbf{v}_1,t)\right] \label{eq:patch}
\end{align}
Thus, our approach to estimating the collision term and the Boltzmann equation
are as follows.  To find the collision {\bf rate} for the region of phase space
associated with a particle $j$, we divide Equation (\ref{eq:patch}) through by
$M_j/m$ (so that we are calculating the scattering probability for a single
macro particle $j$), and we consider only the ``scattering out'' part of the
collision operator.  We consider the pairwise {\bf rate} $\Gamma_{ij}$ for
particle $j$ to scatter off any of the other $i$ particles.  We do a Monte Carlo
simulation of the scatters; if a pair of particles is allowed to scatter in a
given small timestep, we
calculate the macro particles' post-scatter velocity using the center-of-mass
scattering angle $\Omega$.  This latter step is our approximation to the
``scatter out'' term of the Boltzmann collision operator.

The pairwise collision operator is
\begin{align}
  \Gamma_{pq} = \frac{ \Gamma(p|q) + \Gamma(q|p)}{2},
\end{align}
where the conditional probability of scattering a specific particle $p$ off a
target particle $q$ is $\Gamma(q|p)$, which is determined by the collision term
of the Boltzmann equation.  This collision term is derived from Equation
(\ref{eq:patch}), such that
\begin{align}
  \Gamma(p) &= \int_{\delta \mathbf{x}_p} d^3 \mathbf{x} \int_{\delta
\mathbf{v}_p} d^3 \mathbf{v} \int d^3\mathbf{v}_1 \int d\Omega \frac{d
\sigma}{d\Omega} \left| \mathbf{v} - \mathbf{v}_1 \right| (M_p/m)^{-1}
\hat{f}(\mathbf{x},\mathbf{v},t) \hat{f}(\mathbf{x},\mathbf{v}_1,t)
\label{eq:pair1}\\
  &= \int_{\delta \mathbf{x}_p} d^3 \mathbf{x} \int_{\delta \mathbf{v}_p} d^3
\mathbf{v} \int d^3\mathbf{v}_1 \int d\Omega \frac{d \sigma}{d\Omega} \left|
\mathbf{v} - \mathbf{v}_1 \right| (M_p/m)^{-1} \{\sum_j (M_j/m)W(|\mathbf{x} -
\mathbf{x}_j|;h_j)\delta^3(\mathbf{v} - \mathbf{v}_j) \nonumber \\ 
 & \hspace{227 pt} \ \sum_q(M_q/m)W(|\mathbf{x} - \mathbf{x}_q|;
h_q)\delta^3(\mathbf{v}_1 - \mathbf{v_q}) \} \\
  &= \int_{\delta \mathbf{x}_p} d^3 \mathbf{x} \int d^3 \mathbf{v}_1 \int
d\Omega  \frac{d \sigma}{d\Omega} \left| \mathbf{v}_p - \mathbf{v_1} \right|
\sum_q (M_q/m) W(|\mathbf{x} - \mathbf{x}_p|;h_p)W(|\mathbf{x} - \mathbf{x}_q|;
h_q) \delta^3(\mathbf{v}_1 - \mathbf{v}_q) \\
  &= \sum_q \int d\Omega \frac{d\sigma}{d\Omega} \frac{M_q}{m} | \mathbf{v}_q -
\mathbf{v}_p | \int_{\delta \mathbf{x}_p} d^3 \mathbf{x} W(|\mathbf{x} -
\mathbf{x}_p|;h_p)W(|\mathbf{x} - \mathbf{x}_q|; h_q) \\
  &= \sum_q (\sigma/m) M_q | \mathbf{v}_q - \mathbf{v}_p | g_{pq}\\
  &=\sum_q \Gamma(q|p) \label{eq:pair6}.
\end{align}
We note the appearance of the term $\sigma/m$, which is the scattering cross
section per unit mass. The kernel $g$ is defined as 
\begin{align}
 g_{pq} = \int_0^{max(h_p,h_q)} d^3 \mathbf{x}^\prime
W(|\mathbf{x}^\prime|, h_p) W(| \delta \mathbf{x}_{pq} +
\mathbf{x}^\prime|,h_q).
\label{geo.eq}
\end{align}
Using these sets of equations, we calculate
$\Gamma_{pq}$ for each pair of particles whose configuration-space patches
overlap at each time step $\delta t$, making sure to keep the time steps small
enough that
$\Gamma_{pq} \delta t \ll 1$ for each time step.

\section{Test for the Scattering Kinematics}
\label{appendixB}

\begin{figure} \begin {center} 
\includegraphics[width=0.5\textwidth]{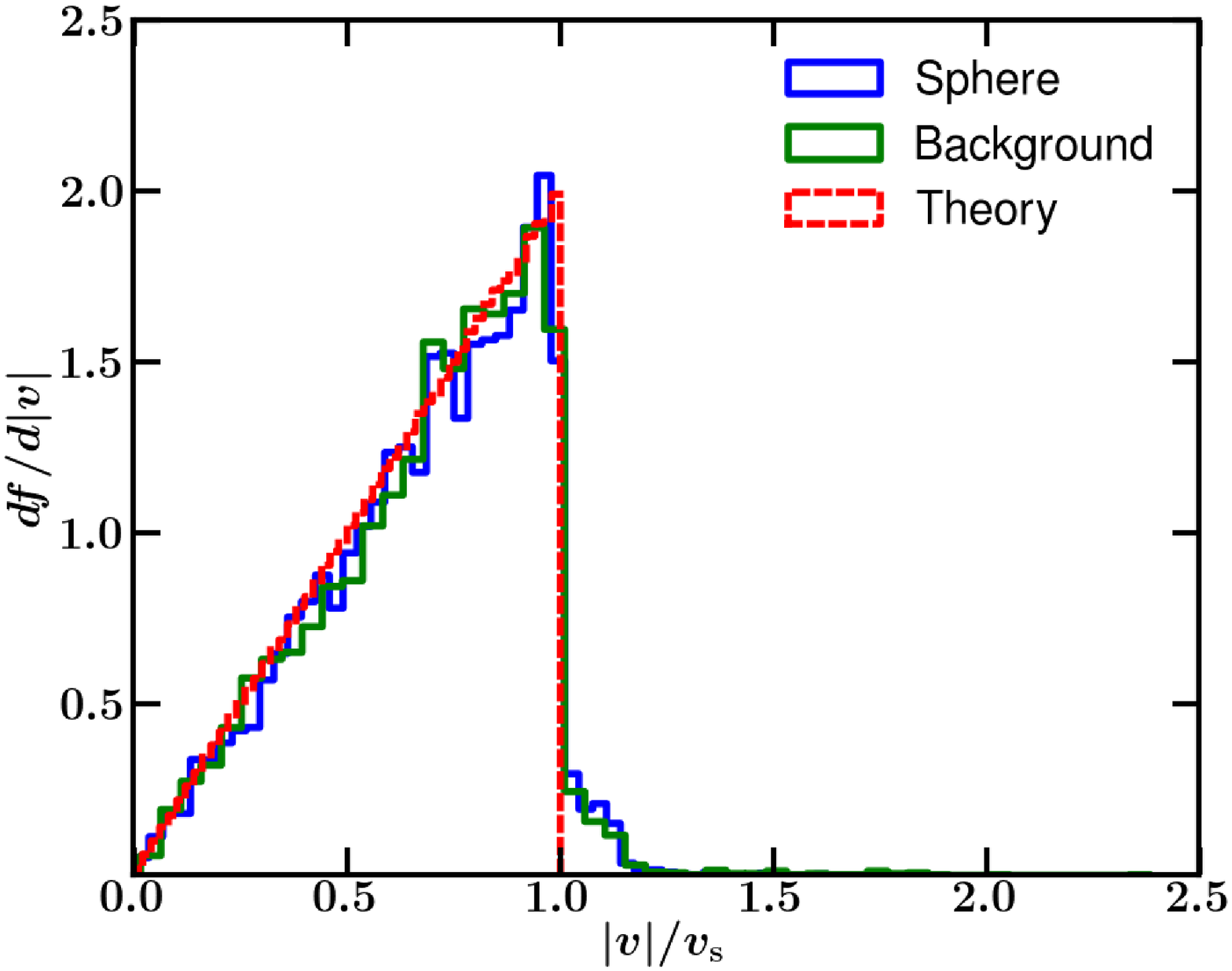}
\end {center} 
\caption{Distribution of the post-scatter velocity magnitudes. From
conservation of energy it is only possible to have particles with velocities
$>v_s$ if they have interacted multiple times, this is not included in our
calculation of the theoretical distribution but it is allowed in our
simulation, hence one can observe a tail for velocities $>v_s$ on
the distributions of both types of particle velocities, but not on the
theoretical distribution.}
\label{vmagDistFig} 
\end{figure}

\begin{figure*} \begin {center} 
\includegraphics[width=\textwidth]{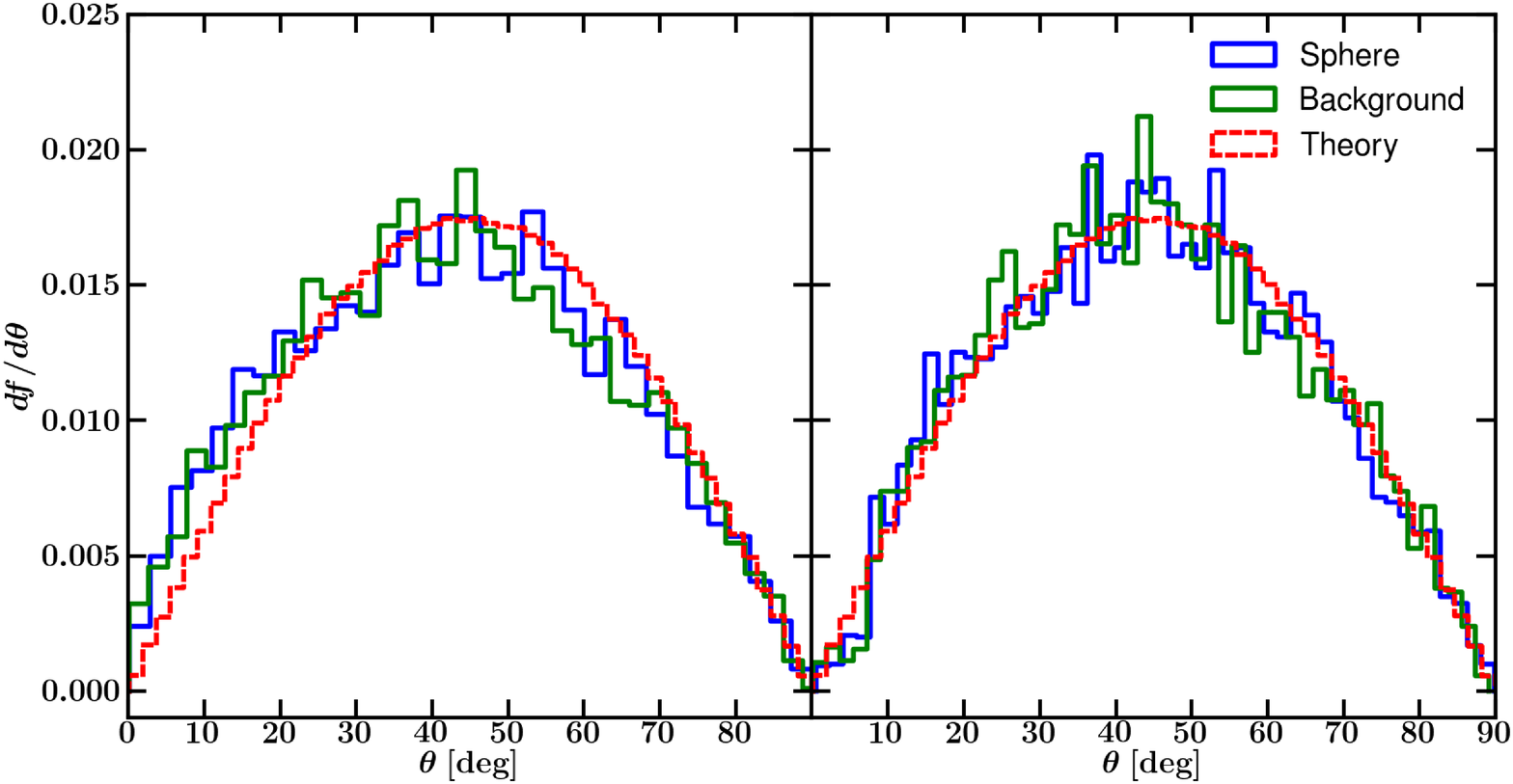}
\end {center} 
\caption{Distributions of the post-scatter velocities along the
$\theta$-directions. It is evident that most of the particles are scattered
towards the $\theta = 45\degrees$ directions, i.e. forming a $45\degrees$ angle
with $\mathbf{v}_s$. Note that the distributions resulting from the
simulation in the left panel are  higher that expected for $\theta < \sim
20\degrees$. This is because multiple scatters are possible in the simulation
and they are not considered in the calculations of the theoretical histograms.
We demonstrate
this by showing in the right panel the distributions from the
simulation when we exclude any particles with $v>v_s$, excluding that way any
particles that we know have interacted multiple times and bringing the
distributions from the simulation to a better agreement with the theory.}
\label{thetaDistFig} 
\end{figure*}

\begin{figure} \begin {center} 
\includegraphics[width=0.5\textwidth]{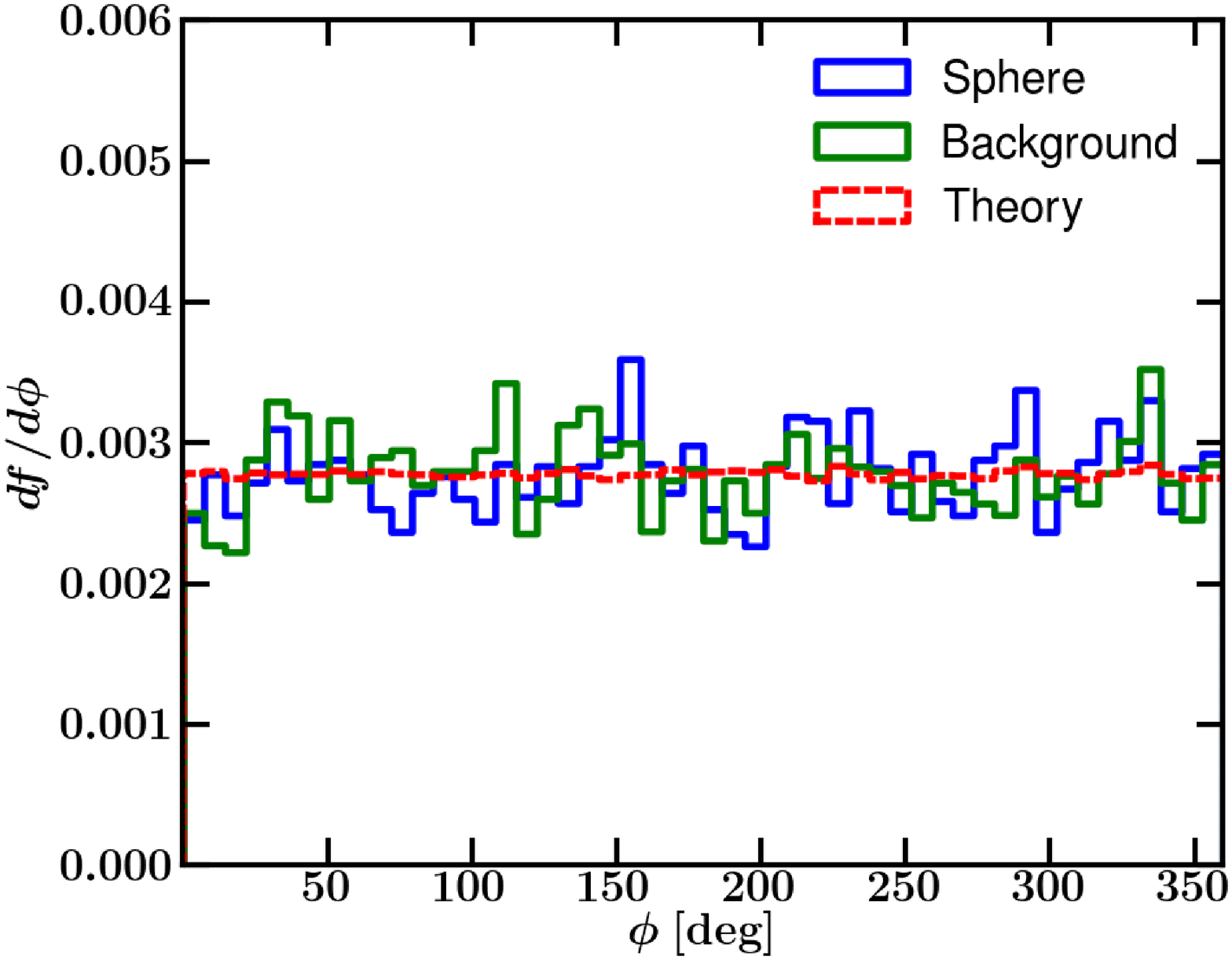}
\end {center} 
\caption{Distributions of the velocities along the $\phi$-directions. The flat
distributions show that the results are symmetric about the direction of motion,
i.e. the z-axis.}
\label{phiDistFig} 
\end{figure}

We use the same setup described in \S \ref{test.sec} to test our
implementation against the expected kinematics. For this we looked at the
distributions of the post-scatter
velocity magnitudes and directions for both the Sphere and Background particles.
For the distributions of the velocity directions we looked at the inclination
and
azimuthal angles of the post-scatter velocity vectors. The angles are defined
such as the line of interaction is along the $\theta=0$ direction (i.e the
z-axis) and $\phi$ is the azimuthal angle about which the experiment is
symmetric. The distributions resulting from our test simulation are
compared to those obtained from the transformation of a uniform
isotropic distribution in the center of mass frame to the simulation/lab frame;
the results are shown in Figures \ref{vmagDistFig}-\ref{phiDistFig}. Figure
\ref{vmagDistFig} shows that the distributions of the velocity magnitudes rise
linearly from $v=0$ to $v=v_s$, followed by a sharp cutoff at $v=v_s$, where
$v_s$ is the relative speed between the sphere and the background. From
conservation of energy it is only possible to have particles with $v>v_s$ if
they have interacted multiple times. Multiple interactions are not considered in
our calculations of the theoretical distributions but they are allowed in our
simulation, hence, in Figure \ref{vmagDistFig} one can observe a tail for
velocities $>v_s$ on the distributions of both types of particle velocities, but
not on the theoretical distribution. Looking at Figure \ref{thetaDistFig}
one can see that most of the particles are scattered towards the $\theta =
45\degrees$ directions, i.e. forming a $45\degrees$ angle with respect
to $\mathbf{v}_s$. It is visible that the distributions resulting from the
simulation in the left panel of Figure \ref{thetaDistFig} are higher than
expected for $\theta < \sim 20 \degrees$. This is again from the
fact that multiple scatters are possible in the simulation and this is 
not included in the calculations of the theoretical histograms. We demonstrate
that this is the case by showing in the right panel of Figure \ref{thetaDistFig}
the distributions obtained from the simulation when we exclude any particles
with $v>v_s$, excluding that way any particles that we know have interacted
multiple times, as one can see doing this brings the distributions from the
simulation to a better match with the theory. Figure \ref{phiDistFig}
shows the distributions of the velocities as a function of $\phi$, these are
flat as expected due to the symmetry of the experiment.

\end{document}